\documentstyle[12pt,epsfig]{article}

\newcommand{\be}{\begin{equation}}
\newcommand{\ee}{\end{equation}}
\def\aprle{\buildrel < \over {_{\sim}}}
\def\aprge{\buildrel > \over {_{\sim}}}
\begin{document}

\topmargin 0pt
\oddsidemargin=-0.4truecm
\evensidemargin=-0.4truecm

\renewcommand{\thefootnote}{\fnsymbol{footnote}}
     
\newpage
\setcounter{page}{1}
\begin{titlepage}     
\vspace*{-2.0cm}
\begin{flushright}
IC/98/43 \\
\vspace*{-0.2cm}
hep-ph/9805272
\end{flushright}
\vspace*{-0.5cm}
\begin{center}
{\Large \bf Parametric resonance of neutrino oscillations and 
passage of solar and atmospheric neutrinos ~~~~~
through the earth}
\vspace{0.5cm}

{\large E. Kh. Akhmedov
\footnote{On leave from National Research Centre Kurchatov Institute, 
Moscow 123182, Russia. E-mail: akhmedov@sissa.it}}\\
{\em The Abdus Salam International Centre for Theoretical Physics,  
I-34100 Trieste, Italy }
\end{center}
\vglue 0.8truecm
\begin{abstract}
We present an exact analytic solution to the neutrino evolution equation 
in matter with periodic step-function density profile and discuss  
in detail the parametric resonance of neutrino oscillations that can
occur in such a system. 
Solar and atmospheric neutrinos traversing the earth pass through 
layers of alternating density and can therefore experience  
parametric resonance of their oscillations. 
Atmospheric neutrinos can undergo parametrically enhanced oscillations
in the earth when their trajectories deviate from the vertical 
by about $26^\circ - 32^\circ$. Solar neutrinos traversing the earth can
experience a strong parametric resonance of their oscillations in a wide 
range of zenith angles. 
If the small mixing angle MSW effect is the
solution of the solar neutrino problem, the oscillations of solar neutrinos 
crossing the core of the earth {\em must} undergo strong parametric 
resonance; this phenomenon should facilitate significantly the observation 
of the day-night effect in oscillations of solar neutrinos. If observed, 
the enhanced day-night effect for core crossing neutrinos would therefore 
confirm both the MSW solution of the solar neutrino problem and the parametric 
resonance of neutrino oscillations. 
\end{abstract}
\end{titlepage}
\renewcommand{\thefootnote}{\arabic{footnote}}
\setcounter{footnote}{0}
\newpage
\section{Introduction}
It is well known that the resonantly enhanced neutrino oscillations 
in matter, the Mikheyev-Smirnov-Wolfenstein (MSW) effect \cite{MS,W}, 
have a simple analogue in classical mechanics: oscillations of two weakly
coupled pendulums with slowly changing frequencies \cite{MS2,Wein}. One 
then naturally wonders if there are any other resonance phenomena in 
mechanics which might have analogues in neutrino physics. 

One such phenomenon is parametric resonance.
The parametric resonance can occur in dynamical systems whose parameters 
vary periodically with time. 
A textbook example is a pendulum with vertically oscillating 
point of support \cite{LL,Ar}. Under certain conditions 
topmost, normally unstable, equilibrium point becomes stable. 
The pendulum can oscillate around this point in the upside-down position. 
Another example, familiar to everybody, is a swing, which is just 
a pendulum with periodically changing effective length. It is the 
parametric resonance that makes it possible to swing on a swing. 

What would be an analogue of the parametric resonance for neutrino 
systems? Since matter affects neutrino oscillations, periodically 
varying conditions can be achieved if a beam of oscillating neutrinos 
propagates through a medium with periodically modulated density. 
For certain relations between the period of density modulation and
oscillation length, the parametric resonance occurs and the oscillations 
get enhanced. The probability of neutrino transition from one flavor state 
to another may become close to unity even for small mixing angle. 

This phenomenon is very different from the MSW effect. Indeed, at the 
MSW resonance the neutrino mixing in matter becomes maximal
($\theta_m=\pi/4$) even if the vacuum mixing angle $\theta_{vac}$ is small.
This leads to large-amplitude neutrino oscillations in a matter of constant
density equal (or almost equal) to the resonance density, or to a strong
flavor conversion in the case of matter density slowly varying along the 
neutrino path and passing through the resonance value. The situation is quite 
different in the case of the parametric resonance. The mixing angle in matter 
(the oscillation amplitude) does not in general 
become large. 
What happens is an amplification of the transition probability 
because of specific phase relationships. Thus, in the case of the 
parametric resonance it is the 
{\em phase} of oscillations (rather than their amplitude) that undergoes 
important modification. 

The parametric resonance of neutrino oscillations has a very simple 
physical interpretation: 
the average value of the transition probability, around which the 
oscillations occur, drifts. This may lead to large probabilities 
of flavor transitions. 
We shall discuss this interpretation in the last section of this paper.  

The possibility of the parametric resonance of neutrino oscillations 
was suggested independently in \cite{ETC} and \cite{Akh}. 
In ref. \cite{ETC} an approximate solution for sinusoidal matter
density profile was found. In \cite{Akh} an exact analytic solution 
for the periodic step-function density profile was obtained. However, in 
ref. \cite{Akh} the results were presented only for a simplified case of 
small matter effects on the oscillations length and mixing angle, 
$l_m\approx l_{vac}$, $\theta_m\approx \theta_{vac}$.  
Parametric effects in neutrino oscillations were further studied in
\cite{KS} where combined action of parametric and MSW resonances and
possible consequences for solar and supernova neutrinos were discussed. 

Recently, there has been an increasing interest in parametric resonance 
of neutrino oscillations. It was pointed out \cite{LS} that 
atmospheric 
neutrinos traversing the earth travel through layers of alternating
density and can therefore undergo parametrically enhanced oscillations. 
The same was also shown to be true for solar neutrinos passing
through the earth \cite{P}. 
Interestingly, this situation 
very closely corresponds to a periodic step-function density profile studied 
in \cite{Akh}. We therefore believe that it would be useful to present the 
exact analytic solution for this case in full, without assuming  
smallness of matter effects. 
We also study the evolution of oscillating neutrinos in the earth
and derive an exact analytic expression for the neutrino evolution 
matrix in the constant-density-layers model of the earth structure. 
Although in this case neutrinos travel only over ``one and a half'' periods 
of density modulation, parametric resonance effects are possible and can 
be very large in this case, too. 
We discuss these effects and their possible manifestations for solar and 
atmospheric neutrinos. 

This paper is organized as follows. The evolution equation for a system 
of oscillating neutrinos in matter is reviewed and the solution for the 
periodic step-function density profile is derived in Sec. 2. Various 
limiting cases are considered and the parametric resonance of neutrino 
oscillations is studied in Sec. 3. Parametric enhancement of
oscillations of solar and atmospheric neutrinos in the matter of the 
earth is considered in Sec. 4. The results are discussed and 
a simple physical interpretation of the parametric resonance in neutrino
oscillations is given Sec. 5. 

\section{Evolution equation and its solution }
Consider a system of two relativistic neutrino species with mixing. 
The evolution of a such a system in matter is described in the weak 
eigenstate basis by the Schr\"{o}dinger equation \cite{W}
\be
i\frac{\partial}{\partial t}
\left(\begin{array}{c}
\nu_a \\
\nu_b
\end{array}\right)=
\left(\begin{array}{cc}
-A(t) & B \\
B & A(t) \end{array} \right)\left(\begin{array}{c}
\nu_a \\
\nu_b
\end{array}\right)\equiv H(t)\left(\begin{array}{c}
\nu_a \\
\nu_b
\end{array}\right)
\label{Sch}
\ee
Here $\nu_{a,b}(t)$ are the probability amplitudes of finding neutrinos 
of the corresponding flavor $a, b$ at a time $t$ (in particular, one of
these two species can be a sterile neutrino $\nu_s$). The parameters $A$
and $B$ 
are 
\be
A(t)=\frac{\Delta m^2}{4E}\cos 2\theta_{vac}-\frac{G_F}{\sqrt{2}}\, N(t)\,,
\quad\quad 
B=\frac{\Delta m^2}{4E}\sin 2\theta_{vac}\,.
\label{AB}
\ee
Here $G_F$ is the Fermi constant, $E$ is neutrino energy, $\Delta 
m^2=m_2^2-m_1^2$, where $m_{1,2}$ are the neutrino mass eigenvalues, and  
$\theta_{vac}$ is the mixing angle in vacuum. The effective density $N(t)$ 
depends on the type of the neutrinos taking part in the oscillations:
\be
N=\left\{\begin{array}{lll}
N_e & {\mbox {\rm for}} &
\nu_e\leftrightarrow \nu_{\mu,\tau} \\
0 & {\mbox {\rm for}} &
\nu_\mu\leftrightarrow \nu_{\tau} \\
N_e-N_n/2 & {\mbox {\rm for}} & 
\nu_e\leftrightarrow \nu_{s} \\
-N_n/2 & {\mbox {\rm for}} &
\nu_{\mu,\tau}\leftrightarrow \nu_{s}\,.
\end{array} \right.
\label{N}
\ee
Here $N_e$ and $N_n$ are the electron and neutron number densities, 
respectively. 
For transitions between antineutrinos one should 
substitute $-N$ for $N$ in eq. (\ref{AB}). If overall matter density
and/or chemical composition varies along the neutrino path, the effective 
density $N$ depends on the neutrino coordinate $t$. The instantaneous
oscillation length $l_m(t)$ and mixing angle $\theta_m(t)$ in matter are 
given by  
\be
l_m(t)=\pi/\omega(t)\,,\quad\quad \sin 2\theta_m(t)=B/\omega(t)\,,
\quad\quad \omega(t)\equiv\sqrt{B^2+A(t)^2}\,.
\label{matter}
\ee
The MSW resonance corresponds to $A(t_{res})=0$, $\sin 2\theta_m(t_{res})=1$.  

We now consider a special case when the effective density $N(t)$ (and 
therefore $A(t)$) is a periodic step function:
\[
N(t)=\left\{\begin{array}{ll}N_1\,, & 0 \le t <T_1 \\
N_2\,, & T_1 \le t <T_1+T_2 
\end{array} \right. 
\]
\be
N(t+T)=N(t)\,,\quad T=T_1+T_2\,.
\label{A}
\ee
Here $N_1$ and $N_2$ are constants. We shall call this the
``castle wall'' density profile. The function $A(t)$ is expressed by 
a similar formula with constants $A_1$ and $A_2$. 
Thus, the Hamiltonian $H(t)$ is
also a periodic function of time with the period $T$: $H(t+T)=H(t)$. 
Let us denote
\be
\delta=\frac{\Delta m ^2}{4E}\,,\quad\quad V_i=\frac{G_F}{\sqrt{2}}\,
N_i\quad\quad (i=1,~2)\,.
\label{not}
\ee
In this notation 
\be
A_i=\cos 2\theta_{vac}\,\delta-V_i\,,\quad\quad 
B=\sin 2\theta_{vac}\,\delta\,,\quad\quad 
\omega_i=\sqrt{(\cos 2\theta_{vac}\,\delta-V_i)^2+
(\sin 2\theta_{vac}\,\delta)^2}\,.
\label{AB1}
\ee

Any instant of time in the evolution of the neutrino system belongs to 
one of the two kinds of the time intervals:
\be
\begin{array}{ll}
(1): & 0+nT \le t <T_1+nT \\
(2): & T_1+nT \le t <T_1+T_2+nT\,, \quad n=0,~1,~2,... 
\end{array} 
\label{int}
\ee
In either of these time intervals the Hamiltonian $H$ is a constant matrix
which we denote $H_1$ and $H_2$, respectively. 
Let us define the evolution matrices for the intervals of time $(0, T_1)$ 
and $(T_1, T_1+T_2)$: 
\be
U_1=\exp(-iH_1 T_1)\,, \quad\quad U_2=\exp(-iH_2 T_2)
\label{U1U2}
\ee
The evolution matrix for the period is then
\be
U_T=U_2 U_1\,.
\label{UT1}
\ee
It is convenient to rewrite the Hamiltonian $H$ given by eq. (\ref{Sch})
using Pauli's $\sigma$ matrices: 
\be
H(t)=B\sigma_1-A(t)\sigma_3\,,
\label{H}
\ee
Let us introduce the unit vectors 
\begin{eqnarray}
{\bf n}_1=\frac{1}{\omega_1}(B,~0,~-A_1)
=(\sin 2\theta_1,~0,~-\cos 2\theta_1)\,, \nonumber \\
{\bf n}_2=\frac{1}{\omega_2}(B,~0,~-A_2)
=(\sin 2\theta_2,~0,~-\cos 2\theta_2)\,,
\label{n1n2}
\end{eqnarray}
where  
$\theta_{1,2}$ are the mixing angles in matter at densities $N_1$ and 
$N_2$: $\theta_1=\theta_m(N_1)$, $\theta_2=\theta_m(N_2)$.  
Then one can write 
\be
H_i=\omega_i (\mbox{\boldmath $\sigma$} {\bf n}_i)\,.
\label{Hi}
\ee
Notice that eigenvalues of $H_i$ are $\pm \omega_i$. 
Using eqs. (\ref{U1U2}) and (\ref{UT1}) 
one arrives at the following expression for the matrix of evolution over 
the period: 
\be
U_T=Y-i\mbox{\boldmath $\sigma$} {\bf X}=
\exp[-i(\mbox{\boldmath $\sigma$}{\bf \hat{X}}) \Phi]\,.
\label{UT2}
\ee
Here 
\be
Y=c_1 c_2-({\bf n_1 n_2}) s_1 s_2\,,
\label{Y}
\ee
\be
{\bf X}=s_1 c_2\,{\bf n}_1+s_2 c_1\,{\bf n}_2-s_1 s_2\,
({\bf n}_1\times {\bf n}_2)\,, 
\label{X}
\ee
\be
\Phi=\arccos Y=\arcsin X\,,\quad \quad \hat{{\bf X}}=\frac{{\bf X}}{X}\,, 
\label{Phi}
\ee
and we have used the notation 
\be
s_i=\sin \phi_i\,,\quad c_i=\cos \phi_i\,,\quad \phi_i=\omega_i T_i\, 
\quad (i=1,~2)\,.
\label{sici}
\ee
Notice that $Y^2+{\bf X}^2=1$ as a consequence of unitarity of $U_T$. 
{}From eq. (\ref{n1n2}) one has
\be
{\bf n}_1 {\bf n}_2=\frac{1}{\omega_1\,\omega_2}(B^2+A_1 A_2)
=\cos(2\theta_1-2\theta_2)\,,
\label{dot}
\ee
\be
({\bf n}_1\times {\bf n}_2)=\frac{1}{\omega_1\,\omega_2} (0,~B(A_2-A_1), 
~0)=(0,~\sin(2\theta_1-2\theta_2),~0)\,.
\label{vector}
\ee
The vector ${\bf X}$ can be written in components as 
\be
{\bf X}=\left(B\left(\frac{s_1 c_2}{\omega_1}+\frac{s_2 c_1}{\omega_2}
\right)\,,~~\frac{B\, s_1 s_2}{\omega_1 \omega_2}(A_1-A_2)\,,~~-\left(
\frac{A_1}{\omega_1} s_1 c_2+\frac{A_2}{\omega_2} s_2 c_1\right)\right)\,.
\label{comp} 
\ee

The evolution matrix for $n$ periods ($n$=1, 2,...) can be obtained by 
raising $U_T$ to the $n$th power:
\be
U_{nT}
\equiv U(t=nT,~0)
=\exp[-i(\mbox{\boldmath $\sigma$}{\bf\hat{X}}) n\Phi] \,.
\label{UnT}
\ee
Eqs. (\ref{UT2})-(\ref{UnT}) give the exact solution of the evolution
equation for any instant of time that is an integer multiple of the period 
$T$. In order to obtain the solution for $nT<t<(n+1)T$ one has to
evolve the solution at $t=nT$ by applying the evolution matrix 
\be 
U_1(t,~nT)=\exp[-iH_1\cdot(t-nT)]
\label{U1t}
\ee
for $nT<t<nT+T_1$ or 
\be
U_2(t,~nT+T_1) U_1=\exp[-iH_2\cdot(t-nT-T_1)] \exp[-iH_1 T_1]
\label{U2t}
\ee
for $nT+T_1 \le t < (n+1)T$, with $H_{1,2}$ given by eq. (\ref{Hi}). 

In the limit $A_2=A_1$ eqs. (\ref{U1U2})-(\ref{U2t}) reproduce the well 
known results for neutrino oscillations in a medium of uniform density, 
as they should. 

\section{Parametric resonance}
Assume that the initial neutrino state at $t=0$ 
is $\nu_a$. The probability of finding $\nu_b$ at a time $t>0$ 
(transition probability) is then $P(\nu_a\to \nu_b,~t)=|U_{21}(t)|^2$
where $U(t)$ is the evolution matrix. 
We will concentrate now on neutrino transition probabilities for $t=nT$.  
Having found it, one can apply the procedure described in the end of the
previous section to find $P(\nu_a\to \nu_b,~t)$ for an arbitrary $t$. 

{}From eq. (\ref{UnT}) one finds 
\be
P(\nu_a\to \nu_b,~t=nT)=\frac{X_1^2+X_2^2}{X_1^2+X_2^2+X_3^2}\,
\sin^2 (n\Phi)=
\frac{X_1^2+X_2^2}{X_1^2+X_2^2+X_3^2}\,
\sin^2 \left( \Phi\frac{t}{T}\right)\,.
\label{prob1}
\ee
This expression resembles the neutrino oscillation probability in a matter 
of constant density. However, 
the pre-sine factor and the argument of the sine 
in (\ref{prob1}) are different from the amplitude and phase of neutrino 
oscillations in matter of either of the two densities, $N_1$ or $N_2$.  
In particular, the pre-sine factor need not be small 
even when both $\sin^2 2\theta_1$ and $\sin^2 2\theta_2$ are small. 
This is a consequence of the parametric enhancement of neutrino
oscillations. The parametric resonance occurs
when the pre-sine factor in (\ref{prob1})  
becomes equal to unity. The resonance condition is therefore     
\be
X_3^2=\left(\frac{A_1}{\omega_1} s_1 c_2+\frac{A_2}{\omega_2} s_2 c_1 
\right)^2=0 \,.
\label{res1}
\ee
For given values of $N_1$, $N_2$, $T_1$ and $T_2$ this equation determines
the resonance energy. At the resonance, $X^2$
takes the value 
\be
(X^2)_{res}=(X_1^2+X_2^2)_{res}=\frac{B^2 s_1^2}{\omega_1^2 \omega_2^2}
\, (A_1-A_2)^2 \left[ 1+c_2^2\frac{B^2}{A_2^2}\right]\,.
\label{Xres1}
\ee
Eqs. (\ref{prob1})-(\ref{Xres1}) give a general description of the parametric 
resonance of neutrino oscillations for the ``castle wall'' density profile
of 
eq. (\ref{A}). 
We shall now analyze the resonance in several particular cases. 

\subsection{Large $|A_i|$}
Assume first that the oscillation frequencies $\omega_{1,2}$ are
dominated by the $A_{1,2}$ terms, i.e. 
\be
A_{i}^2 \gg B^2\,,\quad\quad\quad \omega_{i}\simeq |A_i|\,.
\label{AggB}
\ee
This condition is satisfied in the following cases:

($i$) small vacuum mixing angle and the system is far from the 
MSW resonance, or 

($ii$) $(G_F/\sqrt{2})N_{1,2}\gg \Delta m^2/4E$. 
The condition (\ref{AggB}) is satisfied in this case irrespective of 
the value of $\theta_{vac}$. 

Eq.~(\ref{AggB}) ensures smallness of the oscillation amplitudes in matter 
with effective densities $N_1$ and $N_2$. 
We shall consider two distinct possibilities now. 

{\bf (A) Same sign $A_1$ and $A_2$}. 

\noindent
This corresponds to $N_1,N_2 <N_{MSW}$ or $N_1,N_2 > N_{MSW}$ where 
$N_{MSW}=\cos 2\theta_{vac}\delta (\sqrt{2}/G_F)$ is the MSW resonance
density. In this case 
$X_3^2\simeq \sin^2 (\phi_1+\phi_2)$, and the
parametric resonance condition (\ref{res1}) becomes
\be
\phi_1+\phi_2=\bar{\omega} T=k\pi\,,\quad\quad k=1, 2, ...
\label{res2}
\ee
Here $\bar{\omega}$ is the mean oscillation frequency,
\be
\bar{\omega}=\omega_1 \frac{T_1}{T}+\omega_2 \frac{T_2}{T}\,.
\label{omegabar}
\ee
One can also write the resonance condition (\ref{res2}) as  
\be
\Omega=\frac{2\bar{\omega}}{k}\,, \quad\quad \Omega\equiv
\frac{2\pi}{T}\,.
\label{res3}
\ee
This is nothing but the well known parametric resonance condition in the 
case of small-amplitude oscillations \cite{LL,Ar}. 
Eq. (\ref{res2}) can also be written as 
\be
\bar{l}_m\equiv \frac{\pi}{\bar{\omega}}=\frac{T}{k}\,,
\label{res4}
\ee
i.e. the parametric resonance occurs when the mean oscillation length in
matter $\bar{l}_m$ equals the period of the density modulation divided by an 
integer. 

{\bf (B) Opposite sign $A_1$ and $A_2$}. 

\noindent
This corresponds to $N_1 <N_{MSW}<N_2$ or $N_2< N_{MSW}<N_1$. 
In this case one gets 
$X_3^2\simeq \sin^2 (\phi_1-\phi_2)$, and the resonance condition is 
\be
\phi_1-\phi_2=k'\pi\,,\quad\quad k'=0, \pm 1, ...
\label{res5}
\ee
or 
\be
\left(\omega_0\,\frac{\Delta T}{T}+\frac{\Delta \omega}2\right)=
\frac{k'\,\Omega}{2}\,.
\label{res6}
\ee
Here we have used 
\be
\phi_1-\phi_2=\omega_0\,\Delta T +\frac{1}{2}\,\Delta \omega \,T\,, 
\quad \omega_0\equiv\frac{\omega_1+\omega_2}{2}\,,\quad \Delta T\equiv
T_1-T_2\,,\quad \Delta \omega \equiv 
\omega_1-\omega_2\,.
\label{deltaphi}
\ee

For both cases (A) and (B), at the resonance one has 
\be
(X^2)_{res}=(X_1^2+X_2^2)_{res}=\frac{B^2 s_1^2}{\omega_1^2 \omega_2^2}
\, (A_1-A_2)^2=s_1^2 \,\sin^2 (2\theta_1-2\theta_2)\,.
\label{Xres2}
\ee
This follows from eqs. (\ref{Xres1}) and (\ref{AggB}). In the
vicinity of the resonance $\sin^2 (\phi_1 \pm \phi_2)\simeq [(\phi_1 \pm
\phi_2)-k\pi]^2$, and the transition probability can be written as 
\be
P(\nu_a\to \nu_b,~t=nT)\simeq \frac{D_k^2}{D_k^2+(\omega_{e}-
k\Omega/2)^2}\,
\sin^2\left(
\sqrt{D_k^2+(\omega_{e}-k\Omega/2)^2}
\;t\right)\,. 
\label{prob2}
\ee
Here 
\be
D_k^2=\frac{1}{T^2}\,s_1^2 \,\sin^2 (2\theta_1-2\theta_2)
\simeq \frac{\omega_e^2}{k^2\pi^2}\,s_1^2 \,\sin^2 (2\theta_1-2\theta_2)\,, 
\label{Dk1}
\ee
\be
\omega_{e}=\left\{\begin{array}{lll}
\bar{\omega}=(\omega_1 T_1+\omega_2 T_2)/T\,,& k=1,2,... & {\mbox {\rm 
same sign}} ~
A_1 \;\; {\rm and} ~A_2 \\
\omega_0\,\Delta T/T +\Delta \omega /2\,, & k=0,\pm 1,...
 & {\mbox {\rm opposite  sign}} ~ A_1 \;\; {\rm and} ~A_2 
\end{array} \right.\,,
\label{omegae}
\ee
and we have used the fact that in the case under consideration $X \ll 1$, 
so that $\Phi \simeq X$.

The conditions (\ref{res2}) and (\ref{res5}) determine the sum or the 
difference of the phases $\phi_1$ and $\phi_2$ in the cases (A) and 
(B) respectively, but not the values of the phases themselves. These  
values can be fixed through the following consideration. 
The parametric resonance condition ensures that the pre-sine factor in 
eqs. (\ref{prob1}) and (\ref{prob2}) is equal to unity. This, however, is
not sufficient for the transition probability to be appreciable. 
For this one should also require that the argument of the sine be not too
small. At the resonance, this argument is proportional to $D_k$, so the 
fastest growth of the transition probability is realized when $|D_k|$
reaches its maximum. As follows from (\ref{Dk1}), for fixed values of
mixing angles in matter $\theta_{1,2}$ this amounts to $s_1=\pm 1$, i.e. 
\be
\phi_1=\pi/2 +k'\pi\,.
\label{max}
\ee
Thus, the optimal conditions for having a sizeable transition probability 
are realized when there is a parametric resonance and in addition $D_k^2$
reaches its maximum. This can be summarized as 
\be
\phi_1=\frac{\pi}{2}+k\pi\,, \quad\quad \phi_2=\frac{\pi}{2}+k'\pi\,,
\quad\quad k, k'=0,1,2,...
\label{opt}
\ee
which follows from eqs. (\ref{res2}), (\ref{res5}) and (\ref{max}). The 
conditions (\ref{opt}) apply to both cases (A) and (B). 

The parameter $D_k$ also determines the width of the resonance. The 
pre-sine factor in (\ref{prob2}) becomes equal to 1/2 when the detuning 
$(\omega_e-k\Omega/2)=\pm D_k$. If the parametric resonance condition is 
considered as a condition on neutrino energy, $D_k$
determines the energy width of the resonance. It is easy to find 
the energy width at half height $\Delta E$. The condition (\ref{AggB}) is 
satisfied provided that either $\sin^2 2\theta_{vac}\ll 1$ (barring the 
proximity to the MSW resonance), or $V_i\gg \delta$. In the first
case, as well as in the second case for not too large vacuum mixing angles 
($\cos 2\theta_{vac}\aprge \delta/2 V_i\ll 1$) one finds 
\be
\frac{\Delta E}{E_0}\simeq \frac{2 D_k}{\cos 2\theta_{vac}\,\delta_0} =
\frac{|2 s_1\,\sin(2\theta_1-2\theta_2)|}{\cos 2\theta_{vac}\,\delta_0\,T}
\,.
\label{width1} 
\ee
Here $E_0$ is the resonance energy and $\delta_0=\delta(E_0)$. 
In the case $V_i\gg \delta_0$ and vacuum mixing close to maximal the
result is  
\be
\frac{\Delta E}{E_0}\simeq \frac{2 D_k}{\delta_0}\,\frac{T\, V_1\, V_2}
{\delta_0\,(T_1 V_1+T_2 V_2)}\,,\quad\quad\quad \quad\quad
\cos 2\theta_{vac}\aprle \frac{\delta_0}{2 V_i}\ll 1\,. 
\label{width2}
\ee
The resonance widths $\Delta E/E_0$ in eqs. (\ref{width1}) and
(\ref{width2}) are maximal when $s_1=\pm 1$, i.e. eq. (\ref{max}) is 
satisfied. The widths rapidly decrease with increasing 
resonance order $k$. For the principal resonance, $k=1$ ($k=0$ in the case 
of the opposite sign $A_1$ and $A_2$), 
\be
\frac{\delta E}{E_0}\sim |\sin(2\theta_1-2\theta_2)|\,, ~ 
\delta_0\aprge V_i\,; \quad\quad  
\frac{\delta E}{E_0}\sim |\sin(2\theta_1-2\theta_2)|\,\frac{V_i}
{\delta_0} \,, ~\delta_0\ll V_i \,.
\label{width3}
\ee

Consider now the transition probabilities in more detail. 
If the condition (\ref{max}) is satisfied, $X_{res}=|D_k|_{max}=
|\sin(2\theta_1-2\theta_2)|$, $\Phi_{res}=2(\theta_1-\theta_2)$, and 
the transition probability over one period of density modulation $T$ is 
\be
P(\nu_a\to \nu_b, ~T)=\sin^2 (2\theta_1-2\theta_2)\,. 
\label{prob3}
\ee
The smallness of neutrino mixing in matter, ensured by eq. (\ref{AggB}), 
implies that $\theta_1$ and $\theta_2$ are close to 0 or $\pi/2$,
depending on the signs of $A_1$ and $A_2$. 
We first consider the case (A), when $A_1$ and $A_2$ are of the same sign. 
For $A_1, A_2>0$ ($\theta_{1,2}$ close to 0) the probability (\ref{prob3})
is smaller than the maximal transition probability in matter of constant 
density equal to either $N_1$ or $N_2$, namely, $\sin^2 2\theta_1$ or 
$\sin^2 2\theta_2$. 
When $A_1, A_2 <0$ ($\theta_{1,2}$ close to $\pi/2$) the probability
(\ref{prob3}) is always smaller than the largest of $\sin^2 2\theta_1$ and 
$\sin^2 2\theta_2$. Thus, in this
case the transition probability over the period cannot exceed the maximal
probability in the case of the matter of  
constant density. However, 
the parametric resonance does lead to an important gain. 
In a medium of constant density the transition probability cannot 
exceed $\sin^2 2\theta_m$, no matter how long the distance that neutrinos
travel. On the contrary, in the matter with ``castle wall'' density profile, 
if the optimal parametric resonance conditions (\ref{opt}) are satisfied, 
the transition probability can become large provided neutrinos travel  
large enough distance. For $t=nT$ the transition probability takes the value 
\be
P(\nu_a\to \nu_b,~t=nT)=\sin^2 [2n(\theta_1-\theta_2)] \,.
\label{prob4}
\ee
This probability can become quite sizeable even for small $\theta_1$ and 
$\theta_2$ provided that neutrinos have traveled sufficiently large 
distance. The number of periods neutrinos have to pass in order to 
experience a complete (or almost complete) conversion is
\be
n\simeq \frac{\pi}{4(\theta_1-\theta_2)}\,.
\label{number}
\ee
However, the transition probability can be appreciable even for smaller 
values of $n$. 

Consider now the transition probabilities in the case (B), opposite
sign $A_1$ and $A_2$. Assume for definiteness $A_1>0$, $A_2<0$. 
Since at the MSW resonance $A(t)=0$, in this case $N_1<N_{MSW}<N_2$. 
Small mixing in matter [eq. (\ref{AggB})] then implies that   
$\theta_1$ is close to 0 and $\theta_2$ is close to $\pi/2$. The
transition probability over one period and $n$ periods in the ``optimal
parametric resonance'' are again given by eqs. (\ref{prob3}) and
(\ref{prob4}) respectively. However in this case, for $\theta_2 > \pi/4
+\theta_1/2$ (which is always satisfied for small mixing in matter), one 
has $\sin^2 (2\theta_2-2\theta_1) > \sin^2 2\theta_1,\, \sin^2 2\theta_2$.
This means that {\em even for 
the time interval equal to one period of matter density modulation 
the transition probability exceeds the maximal probabilities of 
oscillations in matter of constant densities $N_1$ and $N_2$.} 
The effect increases with increasing mixing in matter. This is a very
important case; as we shall see in Sec. 4, this parametric enhancement 
is further magnified in the case of neutrinos traveling over ``one and a
half'' periods of density modulation, which has important implications 
for neutrinos traversing the earth. 

\subsection{Small frequency variations}
We will assume here that the frequency variations are small, 
$\omega_1\approx \omega_2$. This is the case when either relative variation 
of density is small, $|N_1-N_2|\ll N_1$, or matter effects are small, 
$V_{1,2} \ll \cos 2\theta_{vac} \,\delta$. 
However, we will not be assuming the condition (\ref{AggB}) in this 
subsection. 

Let us introduce $A_0=(A_1+A_2)/2$ and write 
\be
A_1=A_0+a\,,\quad A_2=A_0-a\,,\quad |a|\ll |A_0|\,.
\label{A1}
\ee 
Then 
\be
\omega_1\simeq \omega_0 \left(1+\frac{A_0\,a}{\omega_0^2}\right)\,, \quad
\omega_2\simeq \omega_0 \left(1-\frac{A_0\,a}{\omega_0^2}\right)\,, \quad
\omega_0 = (\omega_1+\omega_2)/2\simeq \sqrt{B^2+A_0^2}\,.
\label{omegas}
\ee
Notice that the mean oscillation frequency 
\be
\bar{\omega}=
\frac{1}{T}(\phi_1+\phi_2)= \omega_0(1+{\cal O}(\Delta\omega/\omega_0) )\,.
\ee
The resonance condition (\ref{res1}) in this case coincides with 
eq. (\ref{res2}), but eq. (\ref{max}) does not have to satisfied.   

In the vicinity of the parametric resonance $\bar{\omega}\simeq k\Omega/2$ 
one can write 
\be
X_3^2\simeq 
\frac{A_0^2}{\omega_0^2}\,\sin^2(\phi_1+\phi_2)
\simeq
\frac{A_0^2}{\omega_0^2}\,T^2\left(\bar{\omega}-\frac{k\Omega}{2}\right)^2\,,
\label{X3}
\ee
so that the probability of the oscillations becomes 
\be
P(\nu_a\to \nu_b,~t=nT)\simeq \frac{D_k^2}{D_k^2+(\bar{\omega}-k\Omega/2)^2}
\, \sin^2\left(\frac{A_0}{\omega_0}\,\sqrt{D_k^2+(\bar{\omega}-k\Omega/2)^2}
\;t\right)\,. 
\label{prob5}
\ee
The parameter $D_k^2$ here is defined through 
\be
D_k^2\equiv\frac{\omega_0^2}{A_0^2\,T^2}\,(X_1^2+X_2^2)_{res}\,.
\label{Dk2}
\ee
It coincides with the one given in eq. (\ref{Dk1}) in the limit $A_i^2 
\gg B^2$. 
Consider now the case $T_1=T_2=T/2$. We have two distinct situations.

(i) Odd $k$ (this includes the principal resonance, $k=1$). In this case 
$\phi_1-\phi_2\simeq 0$, and 
\be
(X_2)_{res}^2=\frac{B^2(A_2-A_1)^2}{\omega_1^2\,\omega_2}^2
=\sin^2 (2\theta_1-2\theta_2)\,,
\quad\quad\quad
(X_1)_{res}^2\ll  (X_2)_{res}^2\,,
\label{Xres3}
\ee
\be
D_k^2=\frac{B^2}{A_0^2}\,\frac{(A_1-A_2)^2}{\pi^2\, k^2}\,,
\label{Dk3}
\ee

(ii) Even $k$. In this case the main terms in $X_2$ cancel, $X_2={\cal O}
(a^3/A_0^3)$, and $X_{res}$ is dominated by $(X_1)_{res}$ which is of the 
order of $(a^2/A_0^2)$. One finds 
\be
D_k=\frac{1}{4}\,\frac{B(A_1-A_2)^2}{A_0^2}\,.
\label{Dk4}
\ee

For the principal resonance ($k=1$) the resonance width at half-height is 
\be
\frac{\Delta E}{E_0}=\frac{2}{\pi}\,\frac{|A_1-A_2|}{A_1+A_2}\,
\sin 2\theta_{vac}\,\frac{1}{\sqrt{1-[V_0/\omega_0(E_0)]^2\,
\sin^2 2\theta_{vac}}}\,,
\label{width4}
\ee 
where $V_0\equiv (V_1+V_2)/2$. The distance that neutrinos should travel in 
order for the oscillation probability to become equal or close to one is 
$t=nT$ with 
\be
n\simeq \frac{\pi}{2}\,\frac{\omega_0}{A_0}\,\frac{1}{|D_k|\,T}\,.
\label{nn}
\ee
In the limit $V_{1,2}\ll \delta $ the results of ref. \cite{Akh} are
recovered.

\section{Evolution of oscillating neutrinos in the earth}
The earth consists of two main structures -- the mantle, with 
density ranging from 2.7 $g/cm^3$ at the surface to 5.5 $g/cm^3$
at the bottom, and the core, with density ranging from 
9.9 to 12.5 $g/cm^3$ (see, e.g., \cite{Stacey}). The electron number
fraction $Y_e$ is very close to 1/2 both in the mantle and in the core. 
The radius of the earth  $R=6371$ km. At the border of 
the core, which has the radius $r=3486$ km, there is a sharp jump of 
matter density from 5.5 to 9.9 $g/cm^3$, i.e. by about a factor of two. 
This jump is a very important feature of the matter density 
distribution in the earth. The matter densities within the mantle and within 
the core vary little compared to the density jump on the border of the core 
and mantle; therefore to a very good approximation one can consider the 
earth as consisting of mantle and core  of constant densities equal to the
corresponding average densities, $\bar{\rho}_m\simeq 4.5$ $g/cm^3$ and 
$\bar{\rho}_c\simeq 11.5$ $g/cm^3$. Comparison of neutrino transition 
probabilities calculated with such a simplified matter density profile 
with those calculated with actual density profile of the Stacey 
model \cite{Stacey} shows a very good agreement \cite{LS}. More recent 
models of earth give very similar profiles of the matter density
distribution in the earth (for a discussion see, e.g., \cite{BK}).  
We therefore adopt such a ``constant-density-layers'' model of the earth
density distribution 
\footnote{The same approximation was previously used in \cite{Min,Nic}. 
A different analytic approach to propagation of oscillating
neutrinos in the earth was developed in \cite{LiMo}.} .

Interestingly, this model of matter density profile exactly corresponds to 
the periodic step-function (``castle wall'') density profile studied in 
ref. \cite{Akh} and in sections 2 and 3 of this  paper. The peculiarity of
the earth's density profile is that neutrinos traversing the earth do not 
pass through many periods of density modulation; at most, they can only pass 
through, loosely speaking, ``one and a half'' periods  (this would be exactly 
one and a half periods if the distances neutrinos travel in the mantle and 
core were equal). However, as we have seen in the previous section, even in 
the case of one period of density modulation the parametric resonance can 
lead to a very specific amplification of the probability of neutrino 
oscillations; especially for the case of opposite sign $A_1$ and $A_2$ 
the parametric resonance conditions ensure the largest increase of the 
transition probability over one period.  As we shall see, this effect is
further enhanced when neutrinos travel over ``one and a half'' periods of 
density modulation. 

\subsection{Evolution over ``one and a half'' periods} 
Consider now the evolution matrix for neutrinos traversing 
the earth. We will be interested in the situation when neutrinos traverse 
the mantle, core and then again mantle; this happens for the neutrino 
trajectories with the zenith angle $180^\circ \pm 33.17^\circ$ (nadir angle 
$\le 33.17^\circ$).   
The evolution matrix in this case is  
\be
U_3=U_1 U_2 U_1=U_1 U_T=Z-i\mbox{\boldmath $\sigma$}{\bf W}\,,
\label{U3}
\ee
where 
\be
Z=2 c_1 Y-c_2 \,,
\label{Z}
\ee
\be
{\bf W}=2 s_1 Y {\bf n}_1+s_2 {\bf n}_2\,,
\label{W1}
\ee
and $Y$ has been defined in (\ref{Y}). The vector ${\bf W}$ can be written
in components as 
\be
{\bf W}=\left( 2 s_1 Y \frac{B}{\omega_1}+s_2 \frac{B}{\omega_2} \,, ~0\,,
~-\left(2 s_1 Y \frac{A_1}{\omega_1}+ s_2 \frac{A_2}{\omega_2}\right)
\right) \,. 
\label{W2}
\ee
The transition probability in this case is simply 
\be
P(\nu_a\to \nu_b)=W_1^2\,.
\label{prob6}
\ee

Eqs. (\ref{U3})-(\ref{prob6}) give the neutrino transition probability in 
the earth in the constant-density-layers approximation in general case,
i.e. for arbitrary values of oscillation parameters and neutrino energies 
\footnote{The transition probability (\ref{prob6}) applies 
when the initial neutrino state at $t=0$ consists only of $\nu_a$. In
some situations, such as for solar neutrinos, the neutrino state arriving 
at the surface of the earth is a coherent superposition of $\nu_a$ and
$\nu_b$. This case can be easily treated using the evolution matrix
(\ref{U3}). For example, for the probability $P_{2e}\equiv P(\nu_2 \to 
\nu_e)$ relevant for oscillations of solar neutrinos in the earth
\cite{MS3} one finds $P_{2e}=\sin^2 \theta_{vac}+W_1^2 \cos 2\theta_{vac}+
W_1 W_3 \sin 2\theta_{vac} $. 
Different (but equivalent) expressions for $P_{2e}$ were  
derived in \cite{Min} and \cite{P}. For small $\theta_{vac}$ the
probability $P_{2e}$ 
practically coincides with (\ref{prob6}). }. 
We shall now study the conditions for the parametric enhancement of the
oscillations. For fixed values of $N_1$, $N_2$, $\Delta m^2/4E$ and 
$\theta_{vac}$ these conditions give the optimal values of the phases 
$\phi_1$ and $\phi_2$, i.e. of the distances $T_1$ and $T_2$. It is
straightforward to find extrema of $W_1^2$. There are extrema of three types.

(1) Extrema at $c_1=c_2=0$, i.e. $\phi_1=\pi/2+k\pi$, 
$\phi_2=\pi/2+k'\pi$. 
These are exactly the parametric resonance conditions (\ref{opt}) found in
Sec. 3. Thus, in the case of ``one and a half'' periods, as well as in the 
case of $n$ periods, the oscillation probability reaches an extremum   
when the parametric resonance conditions are satisfied. 
The extremum is a maximum provided that either 
\be 
f_1\equiv{\bf n}_1{\bf n}_2-\left (\omega_1/\omega_2-\sqrt{\omega_1^2/
\omega_2^2+8}\right) /4 <0 \,,
\label{f1}
\ee
or 
\be
f_2\equiv {\bf n}_1{\bf n}_2-\left(\omega_1/\omega_2+\sqrt{\omega_1^2/
\omega_2^2+8}\right) /4 >0 \,,
\label{f2}
\ee
otherwise it is a saddle point. 
The functions $f_1$ and $f_2$ are plotted in Fig. 1 along with $A_1$ and 
$A_2$ for $\sin^2 2\theta_{vac}=0.01$. For small $\theta_{vac}$ the 
conditions (\ref{f1}) and (\ref{f2}) are satisfied for $\delta<V_2$ (except 
in a small region of $\delta$ in the vicinity of the point where $A_1=0$).
We shall discuss these conditions in more detail below.

At the resonance the transition probability (\ref{prob6}) takes the value 
\be
P(\nu_a\to \nu_b)=\sin^2 2(\theta_2-2\theta_1)\, \quad\quad(c_1=c_2=0) \,.
\label{prob7}
\ee
The lowest order resonance corresponds to $\phi_1=\phi_2=\pi/2$. 
The parametric resonance of this type is illustrated by Fig. 2. 
The resonance width is rather large; the transition 
probability decreases by a factor of two for 
\be
\Delta\phi_{2,2}=|\phi_{1,2}-\pi/2|\simeq \frac{\pi}{4}\,.
\label{detun}
\ee
Thus, the resonance enhancement of neutrino oscillations can occur 
even for quite sizeable detuning of the phases $\phi_{1,2}$. 

(2) Extremum at $s_1=c_2=0$.  
It is a maximum provided 
$A_2>0$ and a saddle point otherwise. 
The transition probability is 
\be
P(\nu_a\to \nu_b)=\sin^2 2\theta_2\, \quad\quad(s_1=c_2=0)\,.
\label{prob}
\ee
This case corresponds to zero transition probability in the mantle and 
maximal possible transition probability in the core. 
The extremum of this type is illustrated by Fig. 3. It is not a parametric 
resonance. 

(3) Maximum at $c_1^2=s_1^2=1/2$, $s_2=0$, $A_1=0$.  
The latter condition means that the MSW resonance occurs at the density
$N_1$. The transition probability 
is equal to unity. This case is illustrated by Fig. 4; it requires very
special conditions  and we shall not discuss it anymore. 

The behavior of the neutrino transition probability for $\nu_e
\leftrightarrow \nu_{\mu,\tau}$ oscillations as a function of the phases 
$\phi_1$ and $\phi_2$ for various values of $\delta$ is shown in Figs.
2-8. For large $\delta$ for which $A_2>0$ the maximum of the type (2) 
occurs (Fig. 3). In this domain of $\delta$ the conditions (\ref{f1}) and
(\ref{f2}) are not satisfied (see Fig. 1), so the extremum at $c_1=c_2=0$ is 
a saddle point rather than maximum. Eq. (\ref{f1}) is fulfilled in the range 
$9.7\times 10^{-14}~{\rm eV}<\delta<2.12\times 10^{-13}~{\rm eV}$,
which results in a clear parametric peak at $\phi_1=\phi_2=\pi/2$ (Fig. 2).  
For $\delta$ in the range $7.8\times 10^{-14}~{\rm eV}<\delta< 9.7 \times 
10^{-14}~{\rm eV}$ the function $f_1$ is positive while $f_2$ is negative,
so the extremum at $\phi_1=\phi_2=\pi/4$ is again a saddle point; however, 
the value of the transition probability at this point is close to the 
maximal one (Figs. 5 and 6) except in the very close vicinity of the value 
$\delta=\delta_1=8.6\times 10^{-14}$ eV which corresponds to the MSW 
resonance at $N=N_1$. The transition probability for $\delta=\delta_1$ is 
shown in Fig. 4. The maxima of the transition probability are located at 
$c_1^2=s_1^2=1/2$, $s_2=0$, as discussed in the case (3) above. For 
$\delta<7.8\times 10^{-14}$ eV the function $f_2$ becomes positive and the 
extremum at $\phi_1=\phi_2=\pi/4$ again becomes a maximum (Fig. 7). Notice, 
however, that with decreasing $\delta$ the height of the resonance peak 
rapidly decreases. The shape of the transition probability changes especially 
quickly in the region $7.5\times 10^{-14}~{\rm eV}\aprle\delta \aprle  
1\times 10^{-13}~{\rm eV}$, i.e. close to $\delta_1$. For comparison, we
also plotted in Fig. 8 the transition probability for $N_2=N_1$; since matter 
density is constant, no parametric resonance occurs in this case. The
probability of $\nu_{e,\mu,\tau}\leftrightarrow \nu_s$ oscillations
exhibits a similar pattern but for $\delta$ scaled down by a factor of two. 

As follows from (\ref{prob7}), whether or not the parametric resonance can 
lead to a significant increase of the transition probability over the time 
equal to ``one and a half'' periods of density modulation 
depends on the values of the mixing angles in matter $\theta_1$ and 
$\theta_2$. 
The expression (\ref{prob7}) for the transition probability at the
resonance has been found in \cite{LS} and used there to
interpret a specific enhancement of the probability of atmospheric neutrino 
oscillations in the $\nu_\mu\leftrightarrow \nu_s$ channel in terms of the 
parametric resonance. However, in \cite{LS} only the case $\delta\ll V_1, V_2$  
was considered. It is easy to see that, at least for not too large mixing 
in matter, effect is in fact the largest for $\delta$ in the
interval $V_1 <\cos 2\theta_{vac}\delta <V_2$ (opposite sign $A_1$ and
$A_2$), or at least close to this interval. In this range of $\delta$ one has 
$\theta_1=\theta_{mantle}<\pi/4$, $\theta_2=\theta_{core}>\pi/4$, and the 
expression $\theta_2-2\theta_1$ is closer to $\pi/4$ (which gives maximal
transition probability) than either of $\theta_1$ and $\theta_2$ provided 
that $\theta_2>\pi/4+\theta_1$, $\theta_2>3\theta_1$ (small mixing in
matter). Typically, the enhancement effect is even stronger in this situation 
than in the case of oscillations over one period of density modulation.
Indeed, in that case the relevant combination of the mixing angles was 
$\theta_2-\theta_1$ [see (\ref{prob3})]; for $\theta_2>\pi/4+3\theta_1/2$ 
the combination of the mixing angles $\theta_2-2\theta_1$ is closer to
$\pi/4$ than $\theta_2-\theta_1$. 

\subsection{Parametric enhancement of neutrino oscillations in the earth}
In the previous subsection we discussed the behavior of the transition
probability 
as a function of the phases $\phi_1$ and $\phi_2$ treating these phases as 
free parameters. For neutrino oscillations in the earth, however, the phases 
$\phi_1$ and $\phi_2$ are not arbitrary: they depend on neutrino energy 
and on the lengths of neutrino paths in the mantle and core. These lengths 
vary with the zenith angle of the neutrino source, but their changes are 
correlated and they cannot take arbitrary values. It is therefore not obvious 
whether there are any values of the zenith angles and neutrino parameters 
for which the parametric resonance conditions can be satisfied. As we
shall see, the answer to this question is positive.  

Consider neutrinos coming to the detector from the lower hemisphere
and traversing 
the mantle, core and then again mantle. 
As we have mentioned, this corresponds to zenith angles of the neutrino
source $\Theta=180^\circ \pm 33.17^\circ$. The distance $T_1$
that neutrinos travel in the mantle (each layer) and the distance
$T_2$ that they travel in the core are related to the zenith angle
$\Theta$, earth's radius $R$ and core radius $r$ by 
\be
T_1=R
\left(-\cos\Theta-\sqrt{r^2/R^2-\sin^2\Theta}\,\right)\,,
\quad\quad T_2=2R \,\sqrt{r^2/R^2-\sin^2\Theta}\,.
\label{T1T2}
\ee
Assuming the average matter densities in the mantle and core  4.5 and 11.5 
$g/cm^2$ respectively, one finds for the $\nu_e\leftrightarrow \nu_{\mu,\tau}$ 
oscillations involving only active neutrinos the following values of $V_1$ 
and $V_2$: 
\be
V_1=8.58\times 10^{-14}~{\rm eV}\,,\quad\quad\quad
V_2=2.19\times 10^{-13}~{\rm eV}\,.
\label{V1V2}
\ee
For transitions involving sterile neutrinos $\nu_e\leftrightarrow
\nu_{s}$ and $\nu_{\mu,\tau}\leftrightarrow \nu_s$, these parameters 
are a factor of two smaller. 

Consider the lowest-order parametric resonance, 
\be
\omega_1 T_1 =\frac{\pi}{2}\,,\quad\quad\quad 
\omega_2 T_2 =\frac{\pi}{2}\,.
\label{princ}
\ee
Given a value of $\delta$, these conditions yield magnitudes of $T_1$ and
$T_2$ and so fix the zenith angle for which 
they are satisfied (provided, of course, that such an angle exists). 
The problem therefore reduces to finding values of $\delta$ and $\theta_{vac}$ 
for which eqs. (\ref{princ}) have solutions. 
A nontrivial point here is that both conditions
in eq. (\ref{princ}) have to be satisfied at the same value of $\Theta$.  
Notice that these conditions constrain the allowed values of the vacuum 
mixing angle. 
Assuming that the resonance condition can be satisfied 
for all neutrino trajectories that cross the core (including the vertical ones) 
one can derive from the second equality in (\ref{princ}) the following
upper limit 
\footnote{An analogous upper bound following from the first equality  
is less restrictive.}: 
\be
\sin^2 2\theta_{vac}\le \frac{\pi^2}{4(T_2)_{max}^2\,V_2^2}\simeq 0.04\,.
\label{limit}
\ee
If one excludes the zenith angles close to $180^\circ$, the constraint
becomes less stringent. For example, for $\sin^2 \Theta \ge 0.12$ one
obtains $\sin^2 2\theta_{vac}\le 0.07$.  

The analysis shows that the parametric resonance conditions for neutrino 
oscillations in the earth are indeed satisfied in a number of cases; those
include 
$\nu_e\leftrightarrow \nu_{\mu,\tau}$ 
as well as  
$\nu_{e,\mu,\tau} \leftrightarrow \nu_s$ 
oscillations. A systematic study of all the cases fully covering the 
space of parameters $\delta$, $\sin^2 2\theta_{vac}$ and $\Theta$ is 
beyond the scope of the present paper; here we will discuss only a few 
cases of particular interest. In what follows, we shall discuss the 
neutrino trajectories in the earth in terms of the nadir angle 
$\Theta_n=180^\circ-\Theta$.    

($i$) $\nu_e\leftrightarrow \nu_{\mu,\tau}$ oscillations, $\sin^2
2\theta_{vac} \ll 1$ (this case is especially relevant for oscillations of
solar neutrinos in the earth). 
There is a broad range of nadir angles for which 
the parametric resonance conditions (\ref{princ}) are approximately
satisfied: 
$0\le \Theta_n\aprle 
25^\circ$. The maximum of transition probability corresponds to $\delta
\simeq (1.7 - 1.9)\times 10^{-13}$ eV, depending on $\Theta_n$. Fig. 9 shows 
the transition probability (\ref{prob6}) as a function of $\delta$ for 
$\sin^2 2\theta_{vac}=0.01$ and $\Theta_n=11.5^\circ$. There is a parametric
resonance peak at $\delta\simeq 1.9\times 10^{-13}$ eV. A ``shoulder''
on the right slope of the peak is due to the MSW resonance in the core. 
The transition probability at the peak exceeds the biggest of $\sin^2 
2\theta_1$ and $\sin^2 2\theta_2$ (which are the  maximal oscillation 
probabilities in matter of constant densities equal to $N_1$ and $N_2$) by 
about a factor of 3. The values of $\phi_1-\pi/2$ and $\phi_2-\pi/2$ for 
this case are shown in Fig. 10. One can see that these functions vanish at 
very close values of $\delta$, i.e there is a range of $\delta$ (around the 
resonance value $\delta_0\simeq 1.9\times 10^{-13}$ eV) where both conditions 
in eq. (\ref{princ}) are satisfied to a very good accuracy.  
This pattern does not change in the whole interval $0\le \Theta_n\aprle 
25^\circ$: with varying $\Theta_n$ the lowest intersection point of the
curves $\phi_1-\pi/2$ and $\phi_2-\pi/2$ moves slightly along the $\delta$ 
axis but remains close to it. For comparison, we 
have also plotted in Fig. 9 the transition probability for $V_2=V_1$ 
which exhibits no parametric resonance. 

In fact, a strong parametric enhancement of the transition probability
persists up to the nadir angle $\Theta_n=32.9^\circ$ (after which the
resonance peak merges with the MSW resonance at $N=N_1$). The corresponding 
resonance values of $\delta$ are  $\delta_0\simeq (1.1 - 1.5)\times
10^{-13}$ eV. In this region of the nadir angles, the value of
the transition probability at the peak is typically about a factor of 
1.5 -- 2 smaller than it is for $0\le \Theta_n \aprle 25^\circ$. This is 
related to the fact that the conditions (\ref{princ}) are only approximately 
satisfied: the detuning of $\phi_1$ and $\phi_2$ varies between $-0.3$ and
$-0.9$ in this regime. However, since in this case the resonance energy is
farther away from the energies corresponding to the MSW resonance in the 
mantle and core, the relative enhancement of the transition probability is
even stronger: for example, for $\Theta_n=30^\circ$ the transition probability 
at the peak exceeds the biggest of $\sin^2 2\theta_1$, $\sin^2 2\theta_2$
by a factor of 5. 

($ii$) $\nu_e\leftrightarrow \nu_{\mu,\tau}$ oscillations, $\sin^2
2\theta_{vac} = {\cal O}(1)$. This case may be relevant for the large
mixing angle MSW solution of the solar neutrino problem and is also of
interest for atmospheric neutrino oscillations.  For $\sin^2 2\theta_{vac}
\simeq 0.6 - 0.8$ there is a parametric resonance in the narrow range of  
the nadir angles $\Theta_n\simeq 31.1^\circ - 32.7^\circ$. The corresponding
resonance values of $\delta$ are $\delta_0\simeq (6 - 7)\times 10^{-14}$ eV 
(see Figs. 11 and 12). In Fig. 11 a narrower high peak at $\delta\simeq
3.1\times 10^{-13}$ eV is due to the MSW effect in the core.  

($iii$) $\nu_{e,\mu,\tau} \leftrightarrow \nu_s$ oscillations, $\sin^2
2\theta_{vac} \ll 1$. At small nadir angles, $\Theta_n\aprle 18^\circ$, 
the conditions (\ref{princ}) are satisfied for $\delta\simeq (1.5 - 1.7)
\times 10^{-13}$ eV. However, since $\delta>V_2/\cos 2\theta_{vac}$, 
eqs. (\ref{f1}) and (\ref{f2}) are not fulfilled; therefore this 
solution corresponds to a saddle point rather than to a maximum. There is no 
parametric enhancement for this range of $\delta$. 

For a wide range of the nadir angles, $0\le\Theta_n\le 32.6^\circ$, there
is a maximum of transition probability with $\delta_0$ varying from
$1\times 10^{-13}$ eV (at $\Theta_n=0$) to $6.8\times 10^{-14}$ eV
(at $\Theta_n=32.6^\circ$). The corresponding values of $\phi_1$ and
$\phi_2$, though, are $\phi_1\simeq \phi_2\simeq \pi/4$ rather than 
$\pi/2$. 
Nevertheless, this peak can be interpreted 
as a parametric resonance, at least for $\delta_0$ not very close to the
ends of the allowed interval. The point is that the width of the
parametric resonance in the $(\phi_1,\phi_2)$ plane is rather large, and 
detuning $\sim \pi/4$ just decreases the transition probability by about a 
factor of two (see eq. (\ref{detun}) and Fig. 2). Also, the approximate
equality of $\phi_1$ and $\phi_2$ indicates the parametric origin of the peak. 
On the other hand, close to the end points  of the allowed intervals of 
$\delta_0$, i.e. $\delta_0\simeq 6.8\times 10^{-14}$ eV and $\delta_0
\simeq 1 \times 10^{-13}$ eV, the peak values of the transition probabilities 
nearly coincide with either $\sin^2 2\theta_1$ or $\sin^2 2\theta_2$. 
In the vicinity of these points 
an interplay of the parametric and MSW resonances takes place.  

The parametric resonance for $\Theta_n \simeq 30.9^\circ$ ($\delta_0=7.8 
\times 10^{-14}$ eV) is illustrated by Figs. 13 and 14. The transition 
probability at the peak exceeds $\sin^2 2\theta_1\simeq \sin^2 2\theta_2$
by about a factor of 2.7.

($iv$) $\nu_{e,\mu,\tau} \leftrightarrow \nu_s$ oscillations, $\sin^2
2\theta_{vac} = 1$. 
There is a parametric resonance peak for $\Theta_n=25.8^\circ - 32.4^\circ$
and the resonance values $\delta_0=(5 - 7)\times 10^{-14}$ eV. This is the
parametric resonance found in \cite{LS}. 

\section{Discussion}
Neutrino oscillations can be parametrically enhanced in a medium with 
periodically varying density. The periodic step-function (``castle wall'') 
density profile is a very simple example; it allows for an exact 
solution of the neutrino evolution equation and therefore is very 
illuminating. In addition, this example is of practical importance 
since to a good approximation the density profile of the earth can 
be considered as a step-like function with nearly constant densities 
of the mantle and core. 

The parametric resonance realizes very special conditions for neutrino 
oscillations, leading to a possibility of a striking increase of the 
transition probabilities. Typically, neutrinos have to travel over many 
periods of density modulation in order for the parametric enhancement 
to manifest itself. However, under certain conditions even for
neutrino evolution during one period the parametric effects can be quite 
sizeable. Of course, in this case the neutrinos are not exposed to a 
periodic potential, and so one may wonder how the parametric enhancement
of neutrino oscillations can occur. This can be explained as follows. The
parametric resonance implies that the changes of the oscillation phase and 
matter density profile with the coordinate along the neutrino path are 
correlated in a very special way. This ``synchronization'' allows the
transition probability to overbuild after every half-period of density 
modulation (see below); 
if the parameters $A_1$ and $A_2$ defined in eq. (\ref{AB1}) are of
opposite sign, and in addition the oscillation amplitudes at the densities 
$N_1$ and $N_2$ are not too small, a considerable increase of the 
transition probability is possible even for one period. The parametric 
enhancement for one and a half periods of density modulation is then even 
larger. 

It is instructive to examine under what conditions a complete neutrino 
flavor conversion over one period of density modulation is possible. 
Though rather contrived, this case clarifies the essence of the
parametric resonance of neutrino oscillations. 
{}From  eq. (\ref{prob3}) we find that a complete conversion over 
period $T$ would require 
\be
2(\theta_1-\theta_2)=\pm \pi/2\, 
\label{cond1}
\ee
in addition to the conditions (\ref{princ}). 
It is easy to see that, when eq. (\ref{cond1}) is satisfied, 
the transition probability (\ref{prob3}) is equal to the sum of the
amplitudes of the oscillations in matter of constant densities $N_1$ and 
$N_2$, i.e. 
\be
\sin^2 (2\theta_1-2\theta_2)=\sin^2 2\theta_1+\sin^2 2\theta_2\,.
\label{cond2}
\ee
Since at the resonance $\phi_1=\phi_2=\pi/2$, we have the following 
pattern of neutrino oscillations in this case. During the first part of 
the period, $0\le t\le T_1$, usual oscillations in matter of constant
density $N_1$ occur. At the time $t=T_1$ maximal possible in this case 
transition probability is achieved, which is $\sin^2 2\theta_1$. If the 
density remained constant, the transition probability $P$ would start  
decreasing after that and would return to zero at $t=2T_1$. However, 
the density changes to $N_2$ and, as a consequence of the parametric 
resonance, the transition probability continues growing instead of starting 
decreasing. From eqs. (\ref{cond2}) and (\ref{cond1}) it follows that in 
the time interval $(T_1, T_1+T_2)$ the transition probability $P$ undergoes 
one more half-period increase (with the amplitude $\sin^2 2\theta_2$), but
starting from the initial value $\sin^2 2\theta_1$ and not from zero. Thus, 
in this idealized case the
parametric resonance places one half-wave piece of the transition 
probability on the top of the other: the probability never 
decreases until it reaches the maximal value equal to unity. If the 
condition (\ref{princ}) is only approximately satisfied, there is some
decrease of the 
transition probability after the first ``half-period'', but $P$  
does not reach zero, and the decrease is followed by another increase.
What happens is essentially that the average value around which the
transition probability oscillates starts drifting towards larger values. 
In this way $P$ can become quite large even if the amplitude of the 
oscillations around the average value is small. This is illustrated by
Fig. 15, where the dependence of the transition probability  
on the coordinate along the neutrino trajectory 
is shown.   
 
Neutrino oscillations in the earth can undergo a strong parametric 
enhancement.  
Neutrinos coming to the detector from the lower hemisphere from a source 
with the zenith angle in the interval $\Theta=180^\circ \pm 33.17^\circ$
traverse the earth's mantle, core and then again mantle and so pass ``one 
and a half'' periods of density modulation. 
The possibility of parametric enhancement of atmospheric neutrino 
oscillations in the earth was pointed out in \cite{LS} where the
case $\delta=\Delta m^2/4E\ll V_1, V_2$ and the oscillations in the 
$\nu_{\mu}\leftrightarrow \nu_s$ channel were discussed. It was shown 
that the parametric resonance can modify the zenith angle distribution of 
the atmospheric neutrino events.  
As follows from our consideration, the parametric resonance can occur for 
oscillations of atmospheric neutrinos in both $\nu_{\mu}\leftrightarrow 
\nu_{e}$ and $\nu_{\mu}\leftrightarrow \nu_s$ channels for large enough 
nadir angles, $\Theta_n\simeq 26^\circ - 32^\circ$, 

The energy width of the parametric resonance $\Delta E/E_0$ is typically   
$\sim 2 - 3$. Since the atmospheric neutrino anomaly has been observed 
in a rather wide range of neutrino energies (sub-GeV and multi-GeV),   
it seems unlikely that the parametric resonance can alter considerably the 
gross features of the atmospheric neutrino oscillations, such as the
allowed values of neutrino parameters $\Delta m^2$ and $\sin^2 2\theta_{vac}$. 
In any case,  numerical analyses of the atmospheric neutrino data should
have automatically taken the parametric enhancement into account. 
Still, we believe that it may be worth re-analyzing the data paying more 
attention to the parametric resonance and looking for its possible
manifestations.  
It should be also stressed that the parametric resonance can occur in 
oscillations of atmospheric neutrinos even in those channels that are not 
responsible for the atmospheric neutrino anomaly; such effects are potentially 
observable and of considerable interest. 
  
The parametric resonance may be very important for oscillations 
of the solar neutrinos in the earth.  
For small $\sin^2 2\theta_{vac}$, in a wide range of zenith angles
almost completely covering the earth's core, the $\nu_e\leftrightarrow 
\nu_{\mu,\tau}$ oscillations exhibit a strong parametric resonance with the 
peak at $\delta\simeq (1.7 - 1.9)\times 10^{-13}$ eV. 
For typical energy of $^{8}$B solar neutrinos $E\simeq 8$ MeV 
these values of $\delta$ give $\Delta m^2\simeq (5.5 - 6.1)\times 10^{-6}$
eV$^2$. Amazingly, this almost exactly corresponds to the center of the 
allowed interval of $\Delta m^2$ for the so called ``small mixing angle''
MSW solution of the solar neutrino problem (for recent discussions see, 
e.g., \cite{LMP,BK,FoLiMo}). Strong parametric enhancement of the
probability of the oscillations of solar neutrinos in the earth, by up to a 
factor of 7, occurs in the whole allowed region of $\Delta m^2$ and
$\sin^2 2\theta_{vac}$. The small mixing angle MSW solution gives the best 
fit of the available solar neutrino data and so is the most likely solution 
of the solar neutrino problem. The parametric resonance can also occur in the 
$\nu_e\leftrightarrow \nu_s$ oscillations in the earth, but the effect is
smaller in this case. The resonance values $\delta\simeq (6.8 - 10)\times
10^{-14}$ eV correspond to $\Delta m^2\simeq (2.2 - 3.2) 
\times 10^{-6}$ eV$^2$ (for $E=8$ MeV).
These values are close to the lower bound of the allowed interval 
of $\Delta m^2$ for the small mixing angle solution in this channel of
oscillations. For large 
mixing angle the parametric resonance is also possible, but only in a rather 
limited interval of zenith angles of neutrino trajectories. In addition,
the resonance values of $\Delta m^2$ in this case are below the
allowed interval. 

In both cases of $\nu_e\leftrightarrow \nu_{\mu,\tau}$ and $\nu_e
\leftrightarrow \nu_s$ oscillations the parametric effects are strongest
at neutrino energies which are between the energies corresponding to the
MSW resonances in the core and in the mantle of the earth. This is
illustrated by Fig. 16; one can see that the parametric peak at 
$\delta=1.5\times 10^{-13}$ eV is between the maxima of $\sin^2 2\theta_1$ 
and $\sin^2 2\theta_2$ which are due to the MSW resonances in the mantle 
and in the core respectively. 
In other words, the MSW resonance density for neutrinos 
with $\delta\simeq 1.5\times 10^{-13}$ eV is about $7.8~g/cm^3$;  
this density is between the core density and the mantle density. 
The value of the transition probability at the parametric peak exceeds the 
amplitudes of neutrino oscillations in the mantle $\sin^2 2\theta_1$ and in 
the core $\sin^2 2\theta_2$ by more than a factor of six. 

The enhancement of the transition probability in matter in  
the case when the minimal and maximal densities satisfy $N_1<N_{MSW}<N_2$ 
would be of no surprise if the periodic density profile was, e.g., of the 
sinusoidal type. Then somewhere the matter density would be equal
to the MSW resonance density, and one would expect an increase of the 
transition probability \cite{KS}. However, in the case of 
the ``castle wall'' density distribution the MSW resonance condition is 
not satisfied anywhere because the density profile is discontinuous: the
only allowed values of density are $N_1$ and $N_2$. 
It is quite remarkable that in this case, too, there is a sizeable 
enhancement of the transition probability compared to the case of matter 
of constant density. This effect cannot be explained by the usual 
MSW-type increase of the oscillation amplitude; it is related to a 
specific phase relationships  
which are due to the parametric resonance. 

One of the most promising ways of testing the MSW solution of
the solar neutrino problem is to look for the day-night effect due to 
regeneration of solar $\nu_e$'s in the matter of the earth during the  
night \cite{MS3,d-n,numer}. For the most probable small mixing angle
solution, however, this effect is expected to be rather small -- at the level 
of $(1 - 3)\%$. The parametric resonance must increase significantly the 
day-night effect for neutrinos crossing the earth's core. 
Unfortunately, due to their geographical  location, the existing solar
neutrino detectors have a relatively low time during which    
solar neutrinos pass through the core of the earth to reach the detector  
every calendar year. The Super-Kamiokande detector has a largest fractional
core coverage time equal to $7\%$. In \cite{GKR} it was suggested to build
a new detector close to the equator in order to increase the sensitivity 
to the earth regeneration effect; this would also maximize the parametric 
resonance effects in oscillations of solar neutrinos in the earth. 

The parametric enhancement of oscillations of solar neutrinos in the earth 
for neutrinos that travel significant distances in the earth's core 
was discovered numerically (but not recognized as the parametric
resonance) in \cite{numer,Nic,GKR,LMP}. 
The authors of \cite{LMP} found that the day-night effect in the
oscillations of solar neutrinos in the earth can be enhanced by up to a 
factor of six when 
solar neutrinos cross the earth's core. These results were quite
surprising as the largest effect resulted for neutrinos of energies which
correspond to the MSW resonance at the densities of about $(6 - 8)~g/cm^3$, 
the values lying between the core and mantle densities. For neutrinos of 
those energies the MSW resonance is inoperative~ 
\footnote{Strictly speaking, the density profile of the earth is not 
discontinues; the densities $(6 - 8)~g/cm^3$ can be achieved on the border 
of the mantle and core where the density jumps from $5.5$ to $9.9~g/cm^3$. 
However, the jump occurs in a very thin layer 
and so the MSW resonance does not develop.}.  
The possibility of explaining these findings in terms of the parametric
resonance was pointed out in \cite{P}, where also a different interpretation 
was suggested which the author of \cite{P} seems to prefer.
Our results support the parametric--resonance interpretation.

It is remarkable that the parametric resonance conditions can be fulfilled 
for neutrino oscillations in the earth. An important observation 
here is \cite{GKR,LS,LiLu} that the potentials $V_1$ and $V_2$ 
corresponding to the matter densities in the mantle and core, inverse 
radius of the earth $R^{-1}$, and typical values of $\delta\equiv 
\Delta m^2/4E$ of interest for solar (and atmospheric) neutrinos, are all of 
the same order of magnitude -- ($3\times 10^{-14}$ -- $3\times 10^{-13}$) eV. 
It is this surprising coincidence that makes appreciable earth effects on
oscillations of solar and atmospheric neutrinos possible. 
Even more remarkable fact is that the very specific conditions on the
phases of neutrino oscillations in the mantle and core of the earth,
$\phi_1\simeq \phi_2\simeq \pi/2$, which lead to the parametric resonance 
of neutrino oscillations, are also satisfied in many cases of interest. 
Those include atmospheric neutrinos, and, most importantly, solar
neutrinos. 

For small vacuum mixing angle, the parametric resonance occurs for the
values of $\delta=\Delta m^2/4E$ which exactly correspond to the allowed 
values for the small mixing angle MSW effect in the sun.  
If the small mixing angle MSW effect is the solution of the solar neutrino 
problem, the oscillations of solar neutrinos in the earth therefore 
{\em must} undergo strong parametric resonance. 
The parametric resonance provides a natural interpretation for 
the increase of the night-time regeneration effect in the case of
core-crossing neutrinos found numerically in \cite{numer,Nic,GKR,LMP}.  
By selecting only those neutrinos, one can improve the prospects of
observing the day-night effect (though at the expense of reduced 
statistics). If observed, the enhanced day-night effect for core crossing
neutrinos would therefore confirm both the MSW solution of the solar neutrino 
problem and the parametric resonance of neutrino oscillations. 

The author is grateful to Q.Y. Liu and A.Yu. Smirnov for useful
discussions and to S.T. Petcov for informing him about the results of the 
paper \cite{P} prior to its publication. 


\newpage
\centerline{\large Figure captions}

\vglue 0.4cm
\noindent
Fig. 1. Functions $f_1(\delta)$ (upper solid curve) and $f_2(\delta)$ 
(lower solid curve) 
[eqs. (\ref{f1}) and (\ref{f2})]. Also shown are $A_1(\delta) \times
10^{13}$ (dashed line) and $A_2(\delta)\times 10^{13}$ (short-dashed line), 
where $A_{1,2}$ are defined in eq. (\ref{AB1}). $\sin^2 2\theta_{vac}=
0.01$, $\Theta_n=11.5^\circ$.

\noindent
Fig. 2. Transition probability (\ref{prob6}) for $\nu_e\leftrightarrow
\nu_{\mu,\tau}$ oscillations over ``one and a half periods'' of density 
modulation ($t=2T_1+T_2$) vs oscillation phases $\phi_1$ and $\phi_2$ 
acquired during the time intervals $T_1$ and $T_2$. $\sin^2 2\theta_{vac}=
0.01$, $\delta=1.2\times 10^{-13}$ eV.

\noindent
Fig. 3. Same as Fig. 2, but for $\delta=2.5\times 10^{-13}$ eV. 

\noindent
Fig. 4. Same as Fig. 2, but for $\delta=8.6\times 10^{-14}$ eV.





\noindent
Fig. 5. 
Transition probability (\ref{prob6}) for $\nu_e\leftrightarrow
\nu_{\mu,\tau}$ oscillations in the earth for core-crossing
neutrinos vs $\delta$ (solid curve). Also shown is the probability in 
the case $V_2=V_1=8.58\times 10^{-14}$ eV (dashed curve). 
$\sin^2 2\theta_{vac}=0.01$, $\Theta_n=11.5^\circ$.

\noindent
Fig. 6. 
Phases $\phi_1-\pi/2$ (dashed curve) and $\phi_2-\pi/2$ (solid curve) vs
$\delta$ for $\nu_e\leftrightarrow \nu_{\mu,\tau}$ oscillations in the earth. 
$\sin^2 2\theta_{vac}=0.01$, $\Theta_n=11.5^\circ$. 

\noindent
Fig. 7. 
Transition probability (\ref{prob6}) for $\nu_e\leftrightarrow
\nu_{\mu,\tau}$ oscillations in the earth for core-crossing
neutrinos vs $\delta$. 
$\sin^2 2\theta_{vac}=0.6$, $\Theta_n=32.2^\circ$.  

\noindent
Fig. 8. 
Phases $\phi_1-\pi/2$ (solid curve) and $\phi_2-\pi/2$ (dashed curve) vs
$\delta$ for $\nu_e\leftrightarrow \nu_{\mu,\tau}$ oscillations in the earth. 
$\sin^2 2\theta_{vac}=0.6$, $\Theta_n=32.2^\circ$.  

\noindent
Fig. 9. 
Transition probability (\ref{prob6}) for 
$\nu_{e,\mu,\tau}\leftrightarrow \nu_s$ oscillations in the earth for
core-crossing neutrinos vs $\delta$. 
$\sin^2 2\theta_{vac}=0.01$, $\Theta_n=30.9^\circ$.  

\noindent
Fig. 10. 
Phases $\phi_1-\pi/4$ (solid curve) and $\phi_2-\pi/4$ (dashed curve) vs
$\delta$ for $\nu_{e,\mu,\tau}\leftrightarrow \nu_s$ oscillations in the 
earth. $\sin^2 2\theta_{vac}=0.01$, $\Theta_n=30.9^\circ$.

\noindent
Fig. 11. 
Transition probability (\ref{prob6}) for 
$\nu_e\leftrightarrow \nu_{\mu,\tau}$ oscillations 
in the earth as a function of the distance $t$ (measured in units of
the earth's radius $R$) along the neutrino trajectory. $\delta=1.7\times 
10^{-13}$ eV, $\sin^2 2\theta_{vac}=0.01$, $\Theta_n=11.5^\circ$.  

\noindent
Fig. 12. Transition probability (\ref{prob6}) for 
$\nu_e\leftrightarrow \nu_{\mu,\tau}$ oscillations in the earth.  
(solid curve). 
Also shown are $\sin^2 2\theta_1$ (dashed curve) and $\sin^2 2\theta_2$ 
(short-dashed curve). $\sin^2 2\theta_{vac}=0.01$, $\Theta_n=28.6^\circ$.


\newpage
\begin{figure}[H]
\vglue -2cm 
\mbox{\epsfig{figure=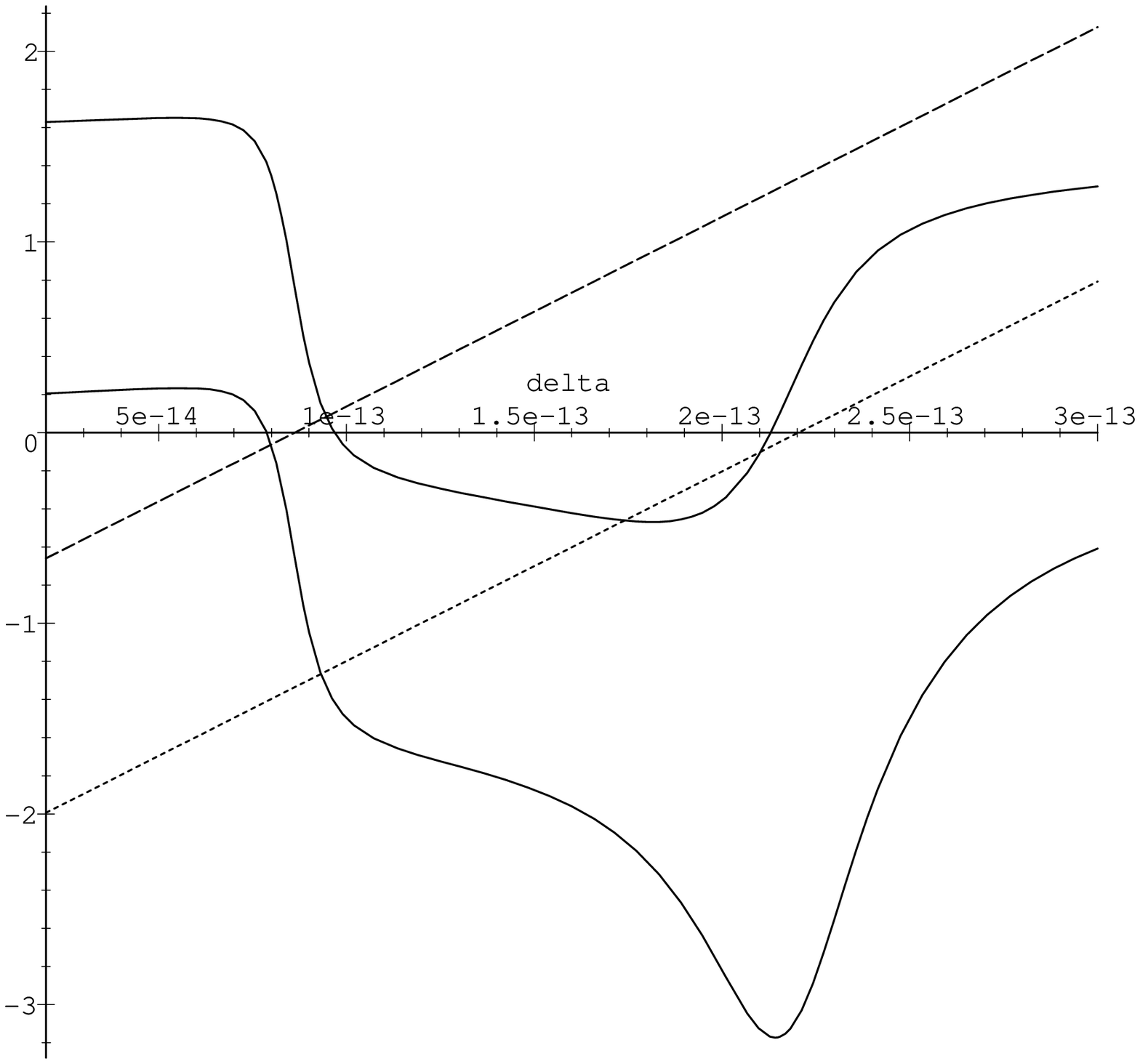,width=15.5cm, height=14cm}}
\vglue -1.5cm
\centerline{\mbox{Fig. 1.}}
\vglue -3cm
\mbox{\epsfig{figure=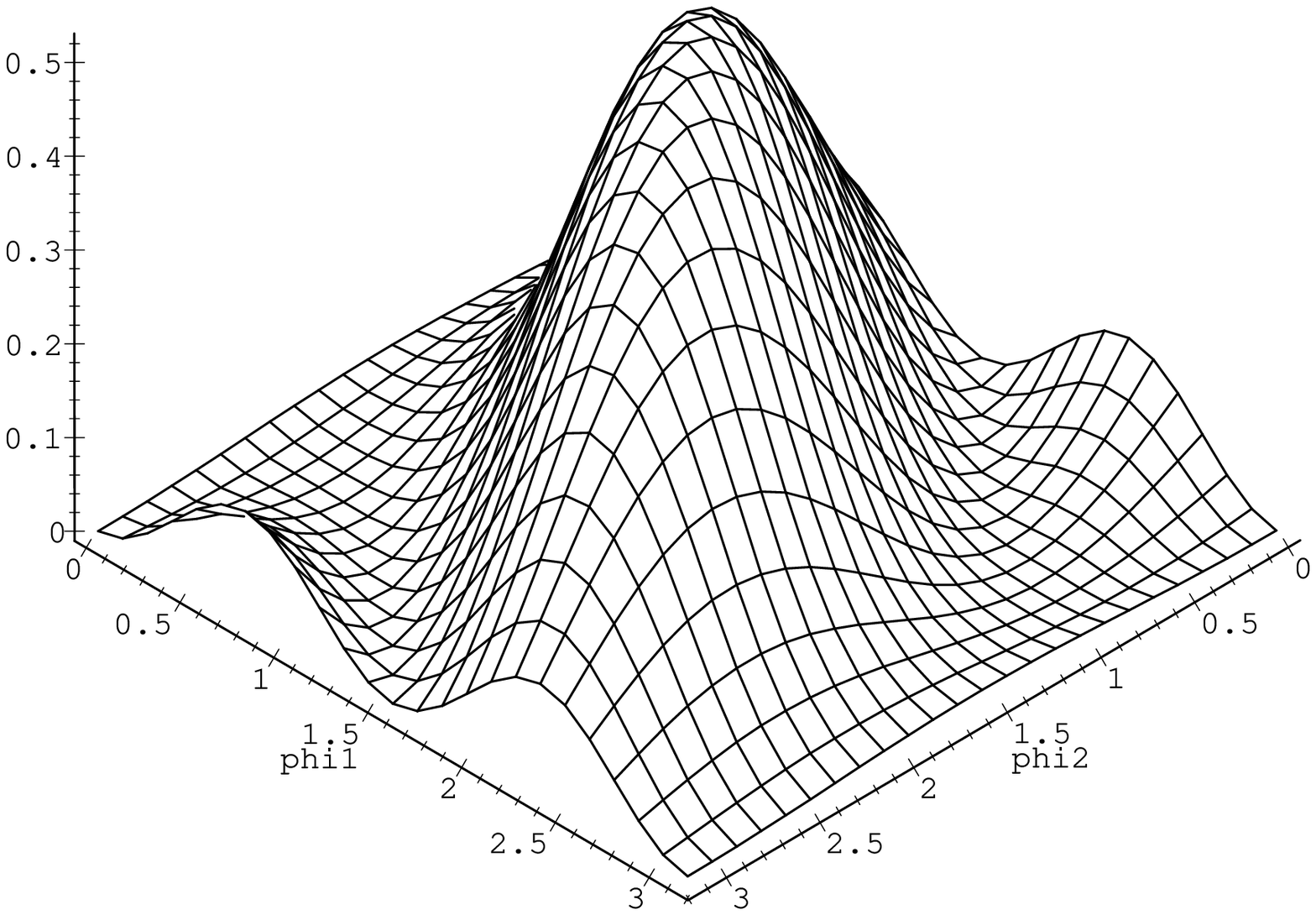,width=15.5cm, height=14cm}}
\vglue -1.5cm
\centerline{\mbox{Fig. 2.}}
\end{figure}

\begin{figure}[H]
\vglue -3cm 
\mbox{\epsfig{figure=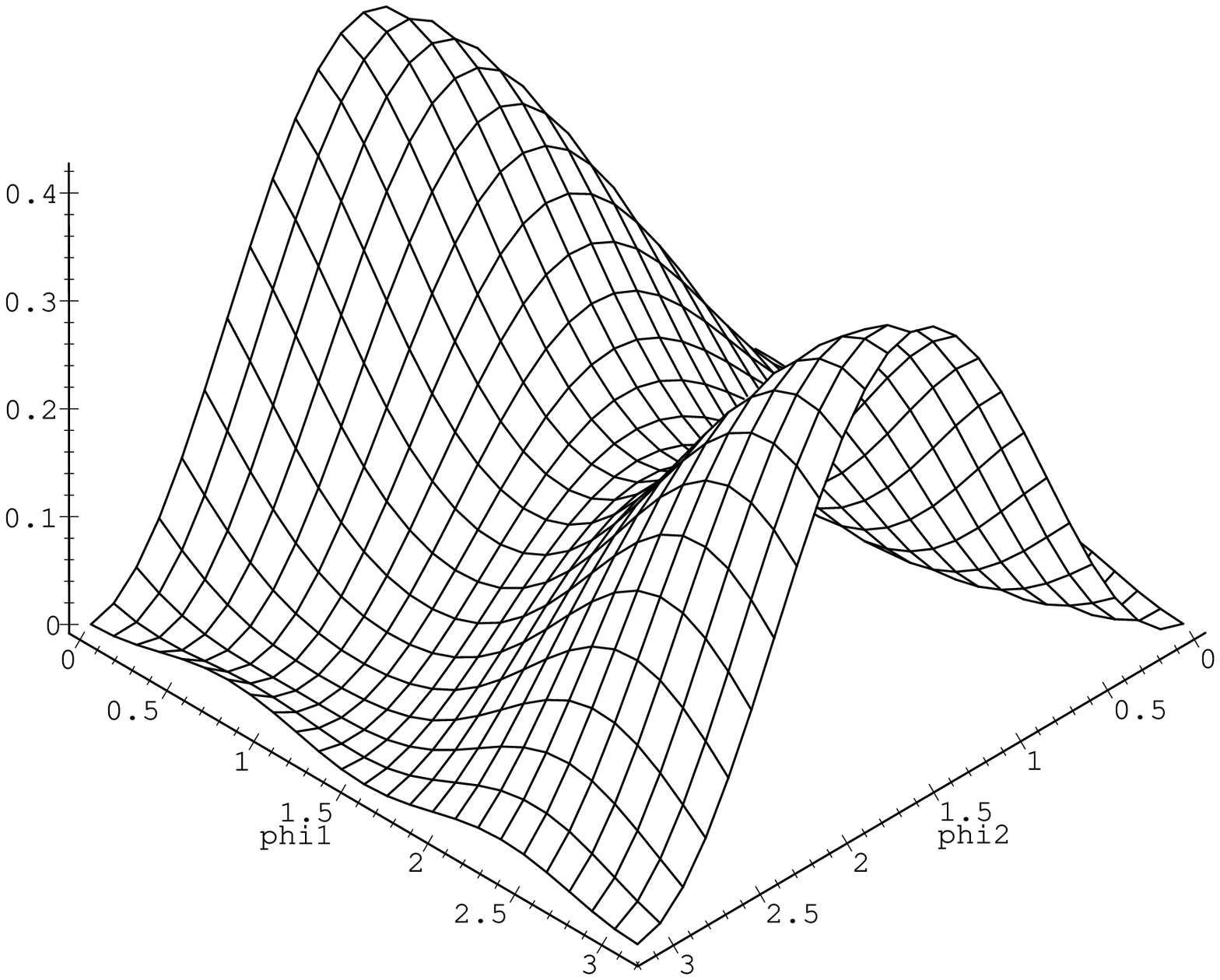,width=15.5cm, height=14cm}}
\vglue -1.5cm
\centerline{\mbox{Fig. 3.}}
\vglue -2cm
\mbox{\epsfig{figure=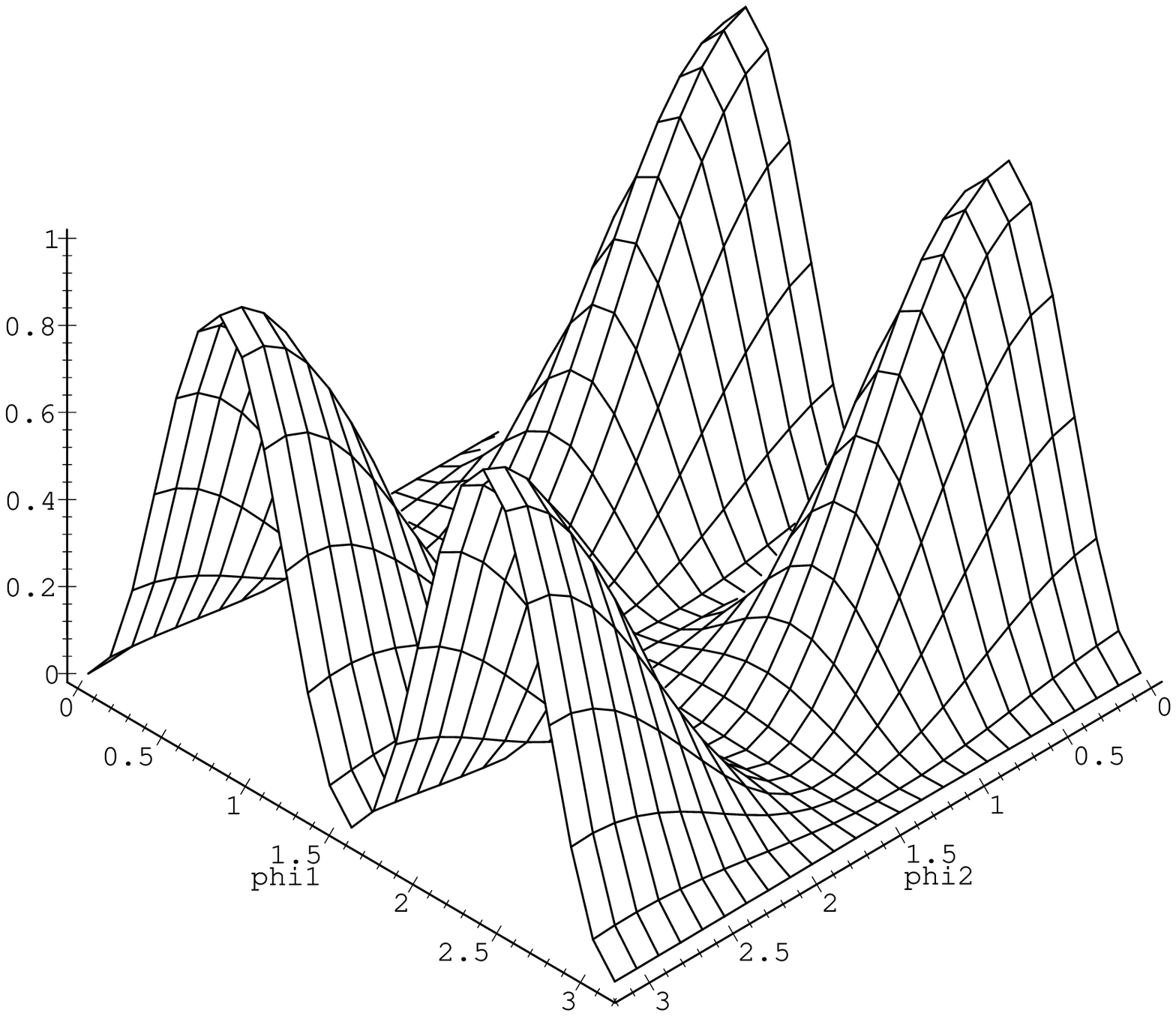,width=15.5cm, height=14cm}}
\vglue -1.5cm
\centerline{\mbox{Fig. 4.}}
\end{figure}



\begin{figure}[H]
\vglue -2cm 
\mbox{\epsfig{figure=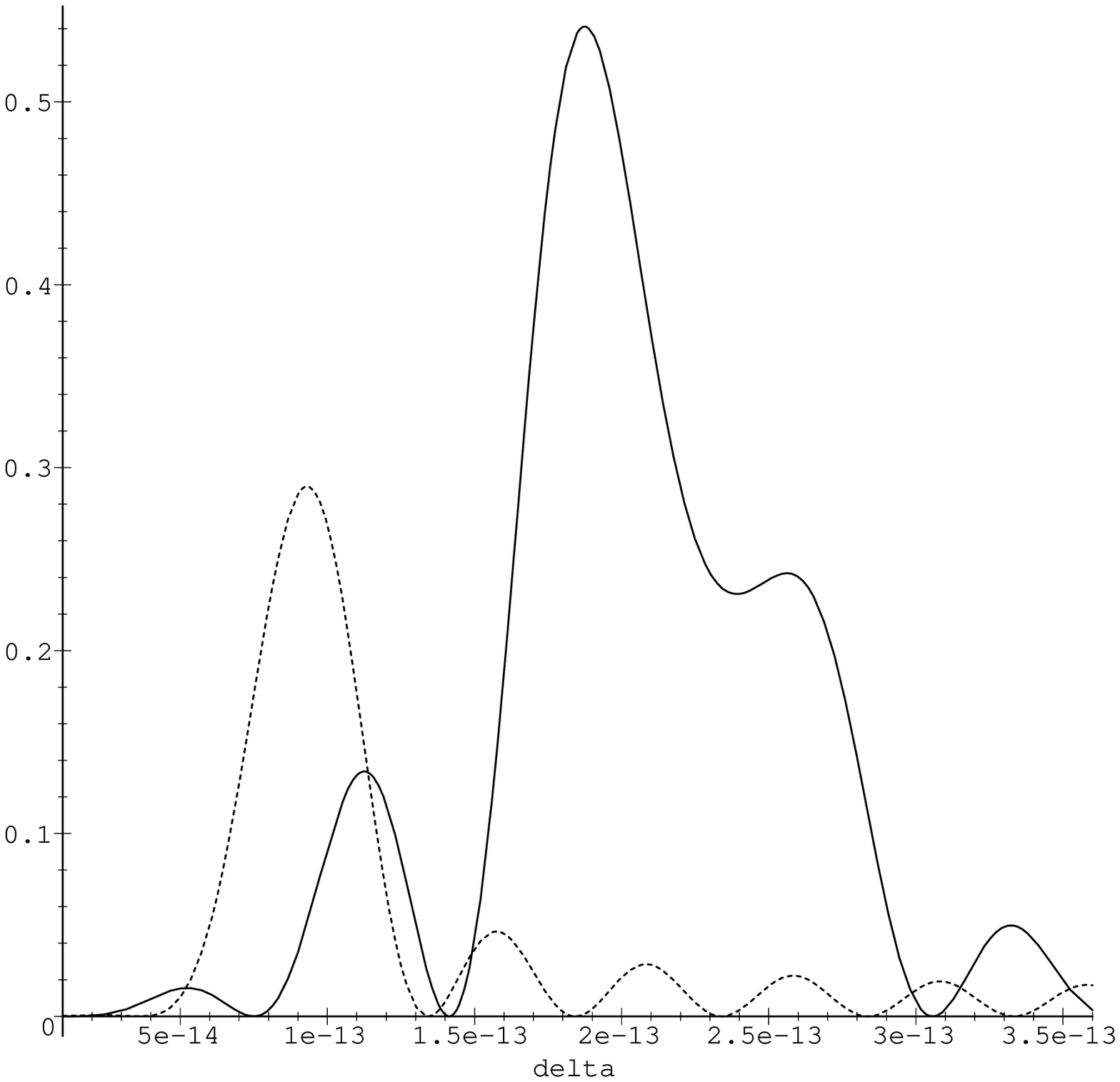,width=15.5cm, height=12cm}}
\vglue -1.0cm
\centerline{\mbox{Fig. 5.}}
\mbox{\epsfig{figure=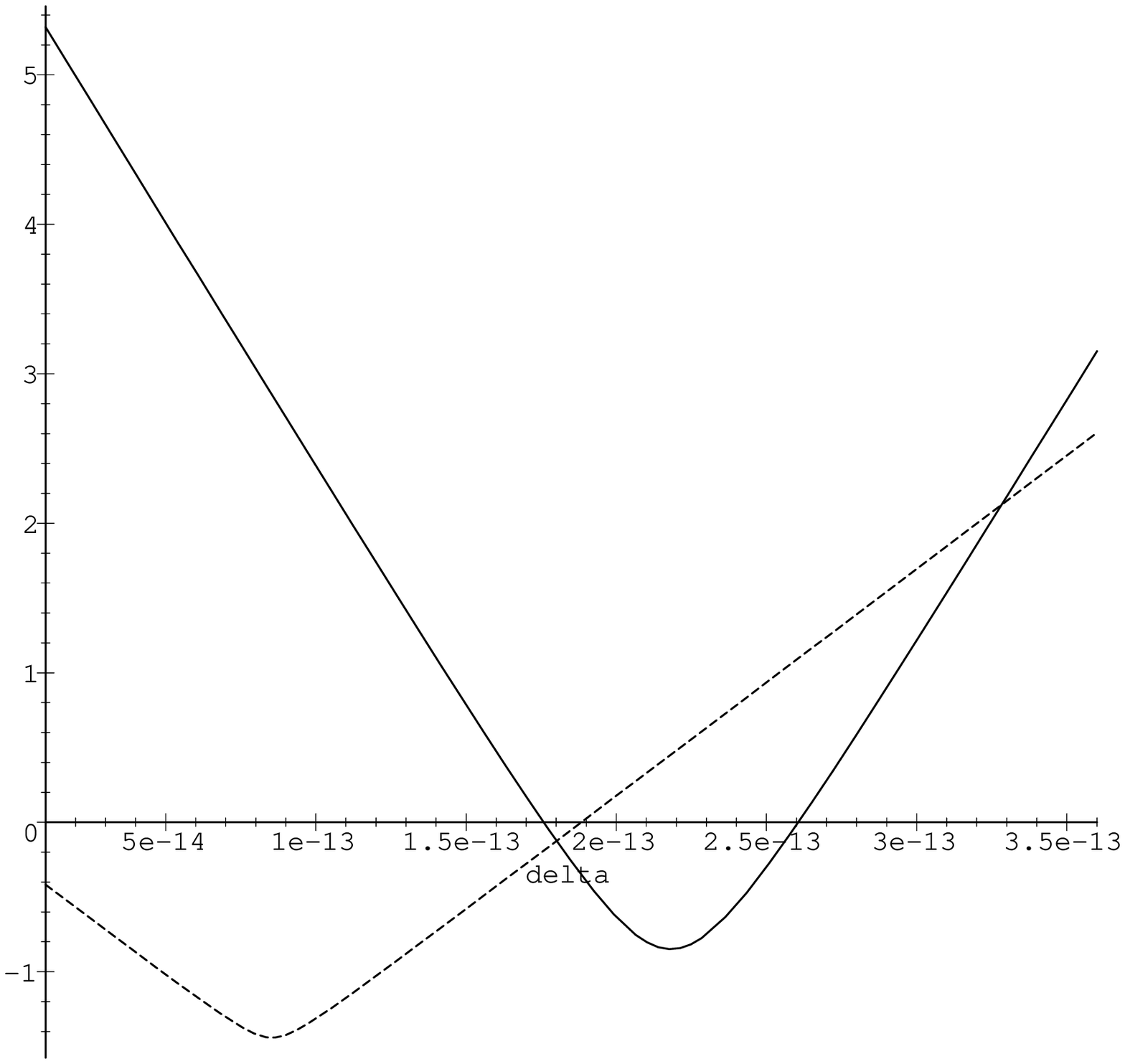,width=15.5cm, height=12cm}}
\vglue -1.0cm
\centerline{\mbox{Fig. 6.}}
\end{figure}

\begin{figure}[H]
\vglue -2cm 
\mbox{\epsfig{figure=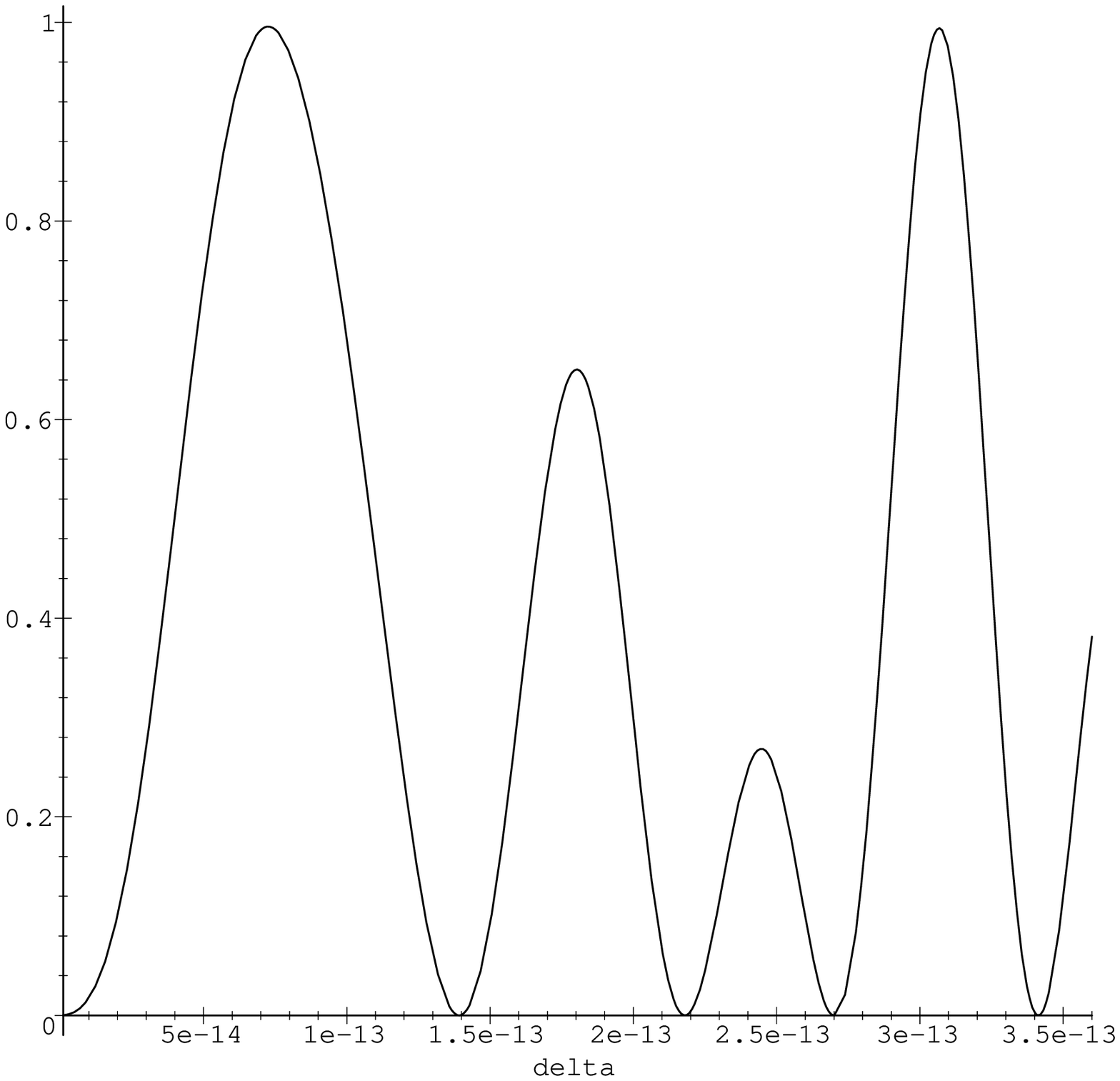,width=15.5cm, height=11.8cm}}
\vglue -1.0cm
\centerline{\mbox{Fig. 7.}}
\mbox{\epsfig{figure=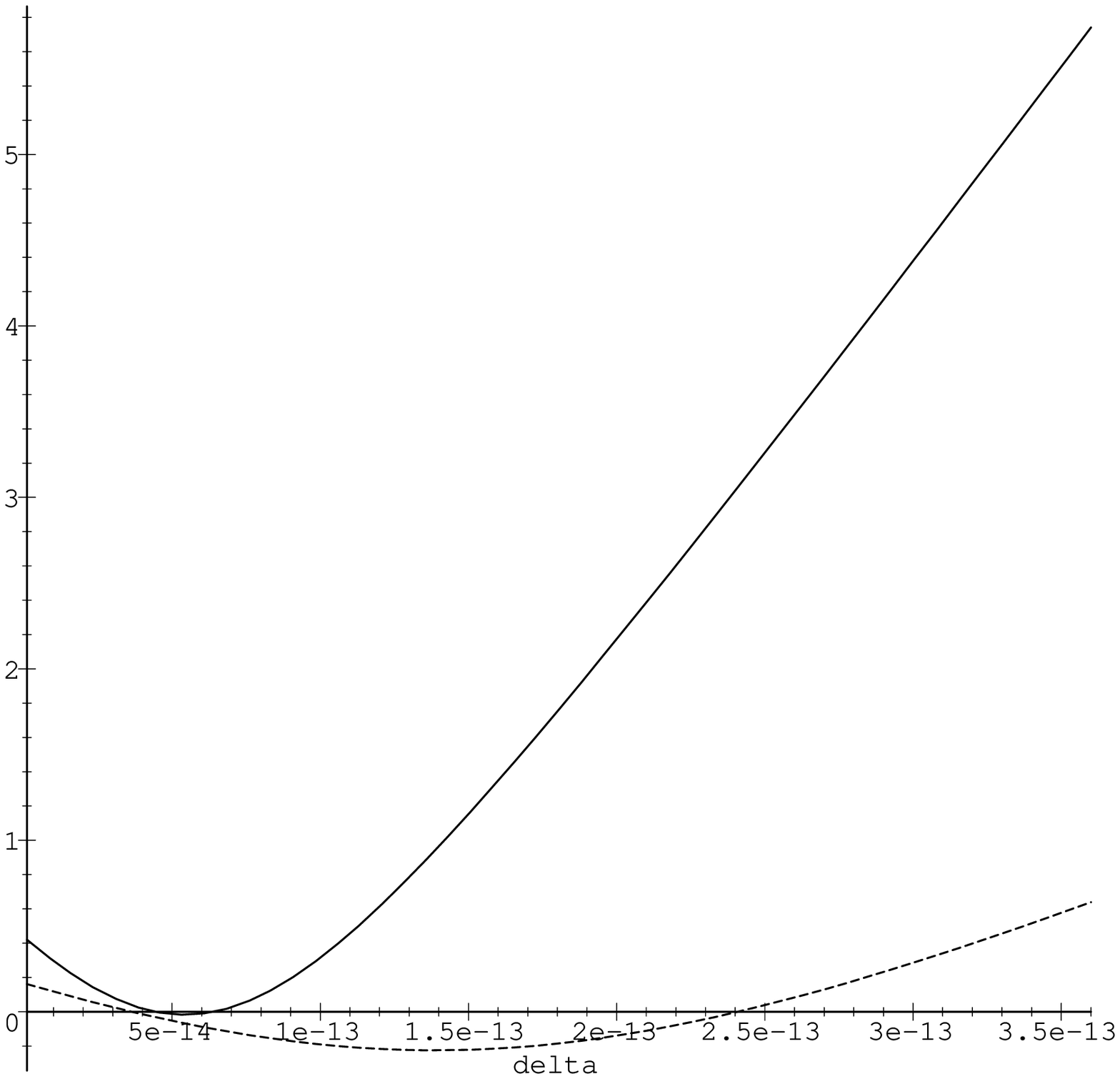,width=15.5cm, height=11.8cm}}
\vglue -1.0cm
\centerline{\mbox{Fig. 8.}}
\end{figure}

\begin{figure}[H]
\vglue -2cm 
\mbox{\epsfig{figure=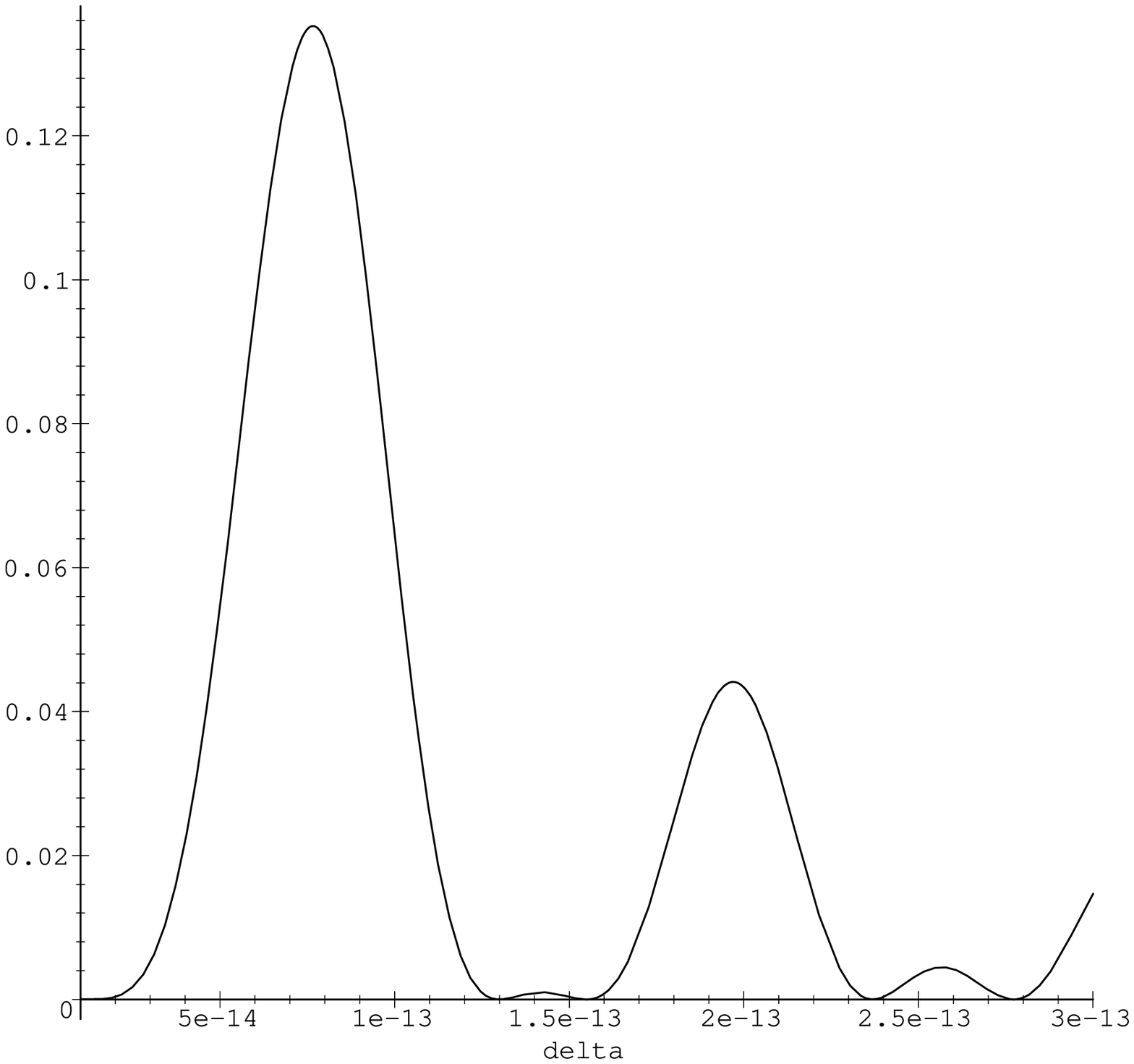,width=15.5cm, height=11.8cm}}
\vglue -1.0cm
\centerline{\mbox{Fig. 9.}}
\mbox{\epsfig{figure=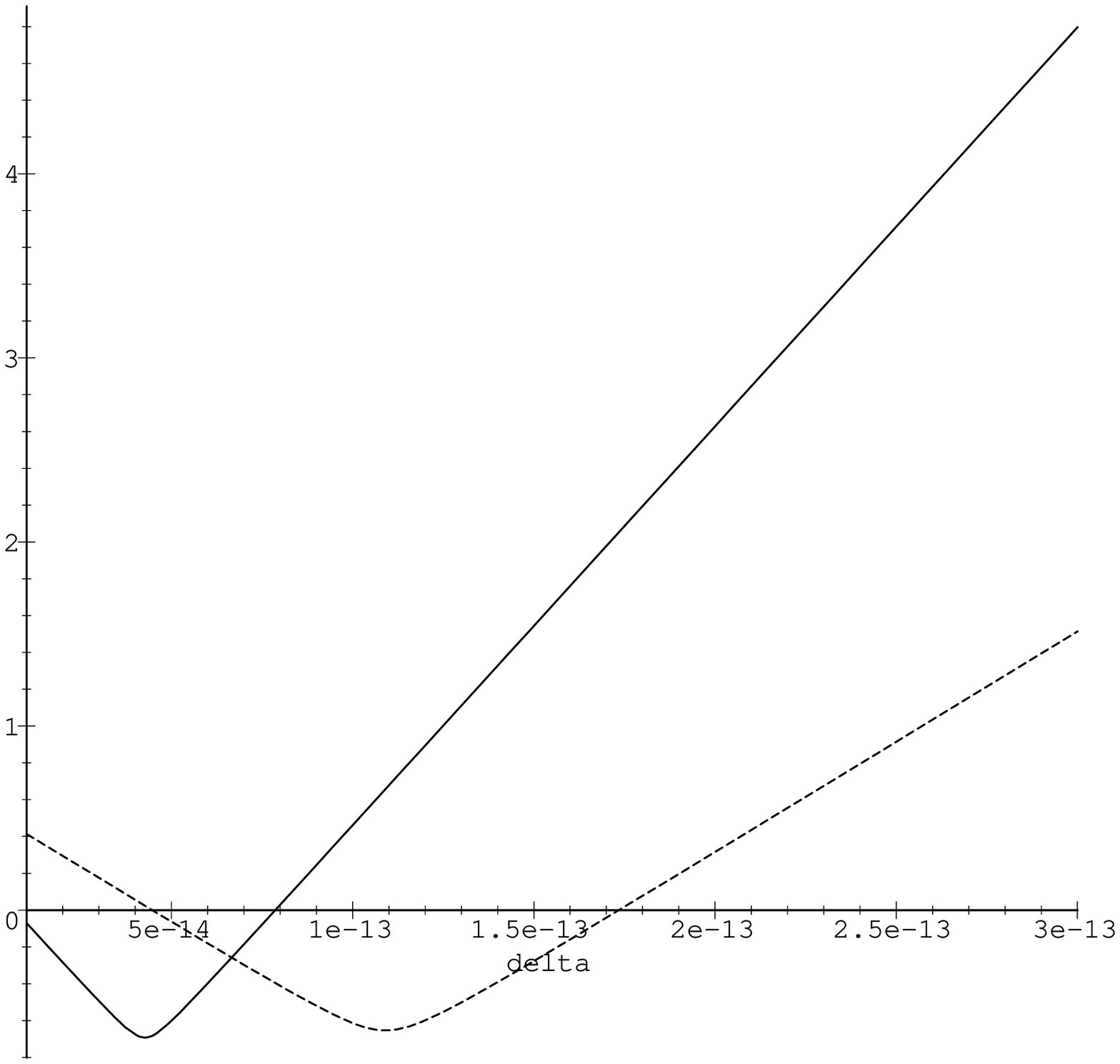,width=15.5cm, height=11.8cm}}
\vglue -1.0cm
\centerline{\mbox{Fig. 10.}}
\end{figure}

\begin{figure}[H]
\vglue -2cm 
\mbox{\epsfig{figure=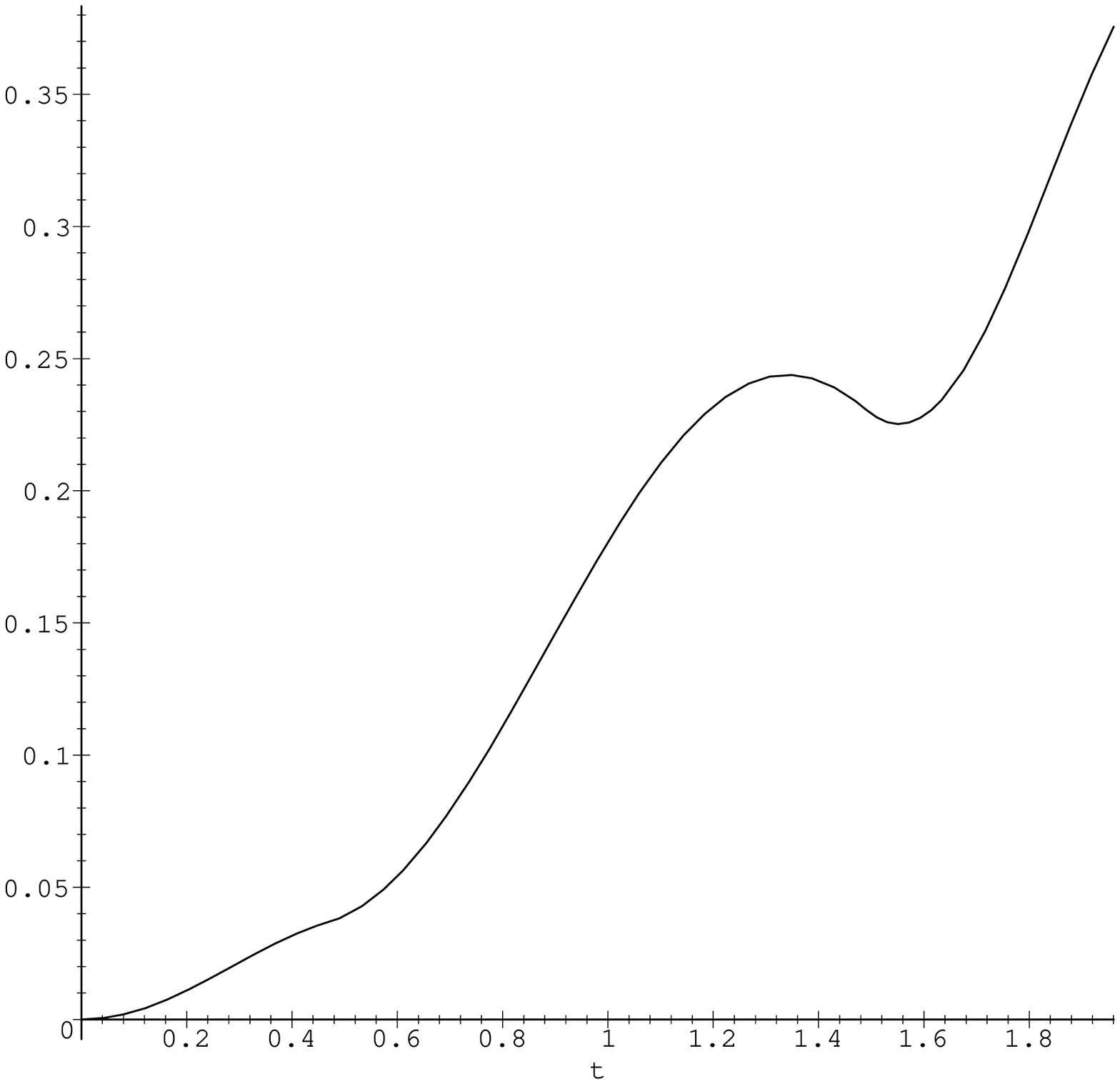,width=15.5cm, height=12cm}}
\vglue -1.0cm
\centerline{\mbox{Fig. 11.}}
\mbox{\epsfig{figure=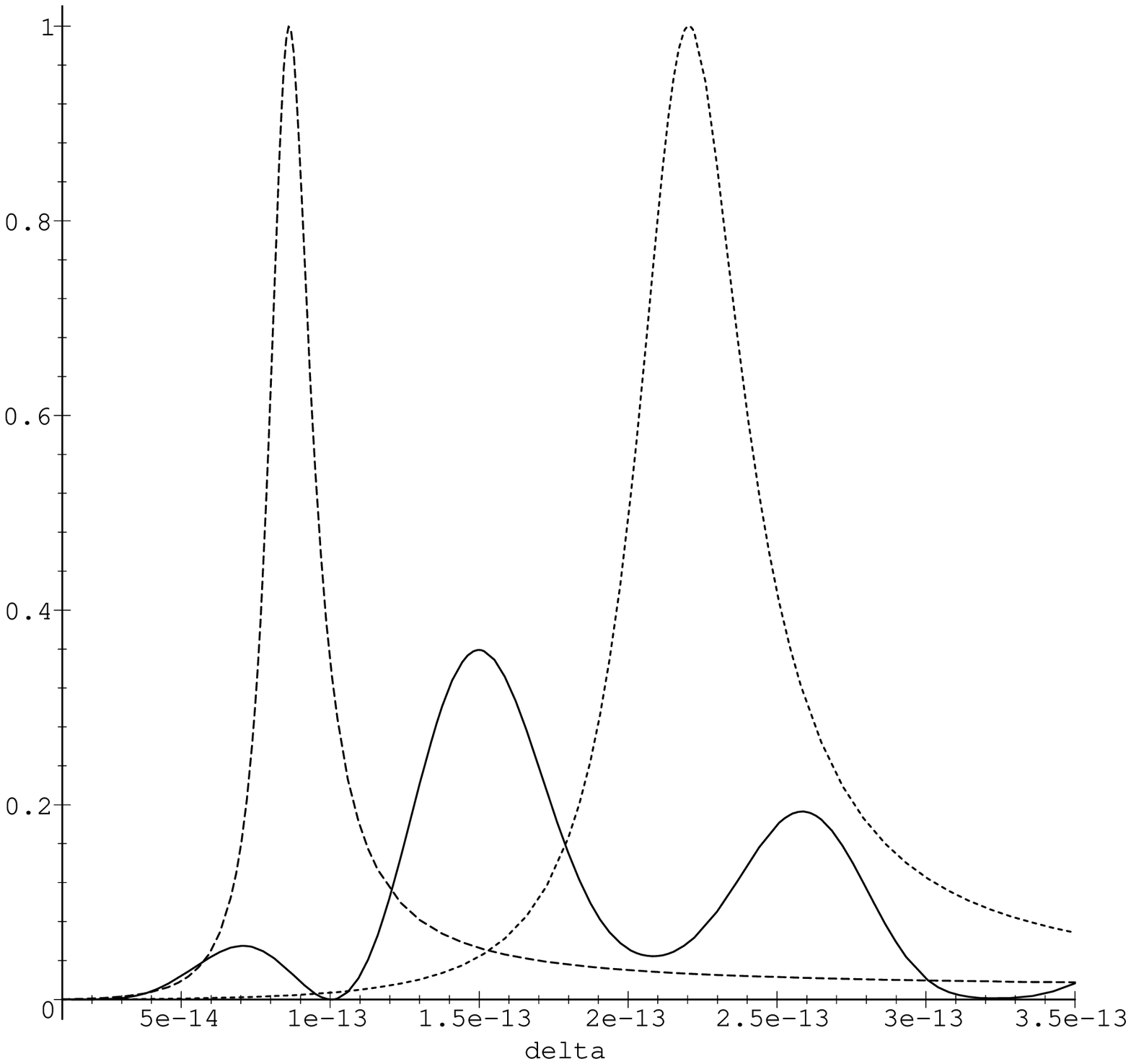,width=15.5cm, height=12cm}}
\vglue -1.0cm
\centerline{\mbox{Fig. 12.}}
\end{figure}

\end{document}